\documentclass[lettersize,journal,twoside]{IEEEtran}
\usepackage{microtype}
\usepackage[normalem]{ulem}
\usepackage{amsmath, mathtools, amsfonts, amssymb, bm}
\usepackage{graphicx}
\usepackage{algorithm, algpseudocode}
\usepackage{siunitx}
\usepackage{bbm, booktabs}
\usepackage[labelformat=simple]{subcaption}

\usepackage{array}
\usepackage{svg}
\usepackage[hidelinks]{hyperref}
\usepackage[capitalize]{cleveref}
\usepackage{cuted}
\usepackage{gensymb} %
\usepackage{dblfloatfix}
\usepackage{multirow}

\DeclareMathOperator*{\argmax}{arg\,max}
\newcommand{\tr}{\operatorname{tr}}
\newcommand{\Frac}[2]{{{#1}/{#2}}}  %

\def\gwtf{g_{\tau,f}^{\rm w}}

\def\taumax{\tau_{\rm max}}

\def\fbA{f_{\rm b, 1}}
\def\fbB{f_{\rm b, 2}}
\def\fbAw{f_{\rm b, 1}^\text{w}}
\def\fbBw{f_{\rm b, 2}^\text{w}}
\def\fbC{f_{{\rm b},i}}
\def\fbCw{f_{{\rm b},i}^\text{w}}

\def\fc{f_{\rm c}}
\def\fs{f_{\rm s}}

\def\taumax{\tau_{\rm max}}
\def\taumin{\tau_{\rm min}}
\def\fmax{f_{\rm max}}
\def\fmin{f_{\rm min}}

\def\xTX{x_{\rm TX}}
\def\xRX{x_{\rm RX}}
\def\xLO{x_{\rm LO}}
\def\xLOQ{x_{\rm LO}^{\rm Q}}
\def\uI{u^{\rm I}}
\def\uQ{u^{\rm Q}}
\def\yI{y^{\rm I}}
\def\yQ{y^{\rm Q}}
\def\etaI{\eta^{\rm I}}
\def\etaQ{\eta^{\rm Q}}
\def\PTX{P_{\rm TX}}
\def\PRX{P_{\rm RX}}
\def\PLO{P_{\rm LO}}
\def\RPD{R_{\rm PD}}
\def\LIFF{\mathcal{L}_{\rm IFF}}

\def\LMF{\mathcal{L}_{\rm MF}}

\def\TwoByTwo{$2 \! \times \! 2$}
\def\SNReta{\mathrm{SNR}_{\eta}}
\def\SNRetaHat{\widehat{\mathrm{SNR}_{\eta}}}

\newcommand*\diff{\mathop{}\!\mathrm{d}}
\newcommand*\dif {\mathop{}\!\mathrm{d}}

\newcommand{\g}[3]{g_{#2,#3}(#1)}

\def\vecu{\mathbf{u}}

\newcommand{\Var}{\mathrm{var}}
\newcommand{\Cov}{\mathrm{cov}}

\newcommand{\delay}{\tau}
\newcommand{\doppler}{f}



\begin{document}

\title{FMCW Lidar Beyond Nyquist by \\ Instantaneous Frequency Fitting}

\author{Alfred~Krister~Ulvog,
Joshua~Rapp,
and~Vivek~K~Goyal%
\thanks{A. K. Ulvog and V. K. Goyal are with Boston University, Boston, MA, 02215 USA (email: keseiya@bu.edu, v.goyal@ieee.org).}%
\thanks{J. Rapp is with Mitsubishi Electric Research Laboratories (MERL), Cambridge, MA, 02139 USA (email: rapp@merl.com). \textit{(Corresponding author: Joshua Rapp)}}%
\thanks{AKU and VKG were supported in part by the US National Science Foundation under Grant 2434818\@.
AKU was supported in part by a gift from Dr.\ John Zheng Sun.
VKG was supported in part by a 2024 Guggenheim Fellowship.
JR was exclusively supported by MERL.}%
}


\maketitle

\begin{abstract}
Frequency-modulated continuous-wave (FMCW) lidar 
conventionally estimates distance and velocity from constant beat frequencies generated through interferometry.
Existing FMCW implementations emphasize simple signal processing---e.g., beat frequency estimation via a fast Fourier transform (FFT) algorithm plus peak-finding---which results in hardware-focused solutions requiring linear swept-frequency laser sources or linearized resampling. 
However, the maximum achievable distance by this method is limited by the need to sample the interference signal  without aliasing.
In this work, we propose 
two signal processing methods: matched filtering and instantaneous frequency fitting.
These two methods can recover larger ranges of distance and velocity by considering the full waveform despite aliasing in the frequency domain. Furthermore, the FMCW lidar signal is often corrupted by phase noise, and we show that the instantaneous frequency fitting approach is more robust than matched filtering by considering the deviation in the phase. 
We present comprehensive simulation studies along with theoretical analysis using the misspecified Cram\'er--Rao bound.
As these methods are flexible to arbitrary frequency modulation, we also show results for non-linear modulations that could yield better sensitivity to distance and velocity compared to the popular triangular modulation.

\end{abstract}


\section{Introduction}

Frequency-modulated continuous-wave (FMCW) lidar uses interferometry to produce high-resolution distance measurements~\cite{uttam_precision_1985, kubota_interferometer_1987}.
An increasing number of commercial products now combine FMCW lidar with beam-steering to achieve dense 3D point clouds 
for
applications such as
autonomous navigation~\cite{li_lidar_2020, roriz_automotive_2022}.
Compared to incoherent (time-of-flight) lidar, FMCW lidar is more robust to ambient noise and interference and requires relatively lower sampling rates~\cite{behroozpour_lidar_2017}.
The ability to configure FMCW lidar for simultaneous distance and radial velocity measurement---sometimes branded as ``4D lidar''~\cite{lin_4-d_2023}---is also considered a major advantage for robotics and navigation applications~\cite{pierrottet_navigation_2011, watts_lidar_2023, piggott_coherent_2025},
and large-scale focal plane arrays for 4D coherent imaging have recently been unveiled~\cite{settembrini_large-scale_2026}.

\begin{figure*}
\centering
{\phantomsubcaption\label{fig:first_a}}
{\phantomsubcaption\label{fig:first_b}}
{\phantomsubcaption\label{fig:first_c}}
{\phantomsubcaption\label{fig:first_d}}
\includegraphics[trim={0 21mm 0 20mm}, clip, width=\linewidth]{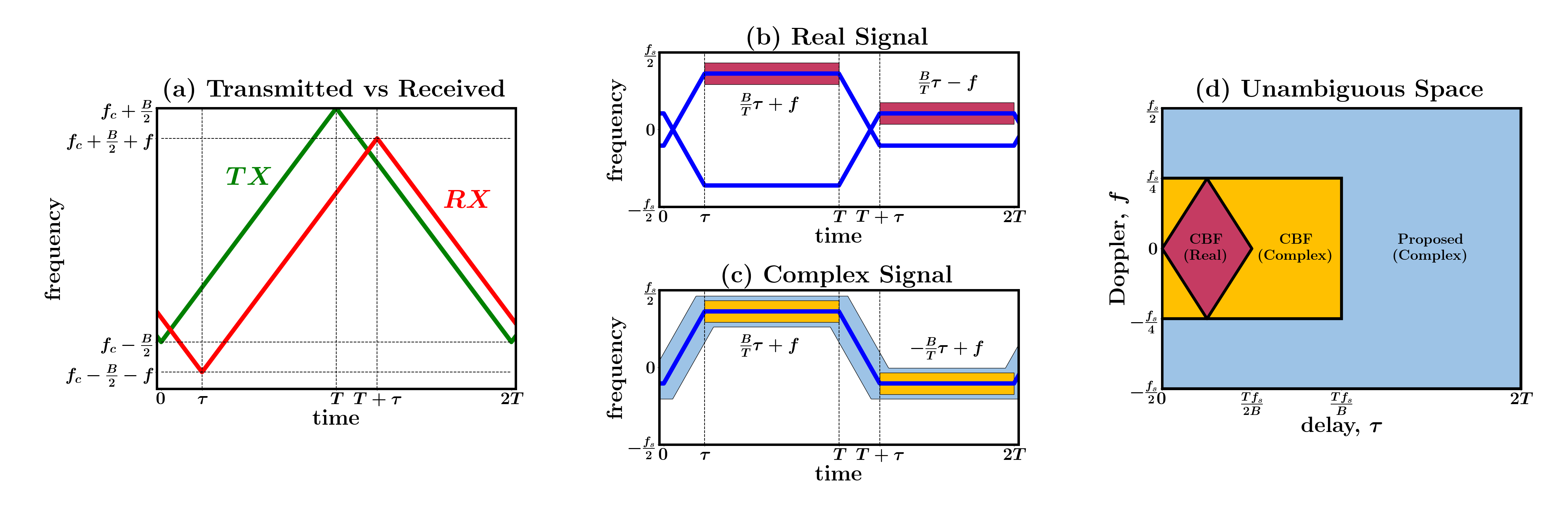}
\caption{
Overview of conventional FMCW lidar processing and our proposed approach. 
(a) FMCW lidar interferes frequency-modulated transmitted (TX) and received (RX) signals. 
When the transmitted signal is piecewise linear, conventional methods estimate distance and velocity from constant beat frequency (CBF) regions in either real-valued (b) or complex-valued (c) interference signals.
Our proposed method leverages the entire signal, including the non-CBFs regions, and thus achieves a much larger unambiguous space of distance and velocity (d). 
Region colors in (d) correspond to the shaded signal segments in (b) and (c).
}
\label{fig:firstfig}
\end{figure*}

In FMCW lidar, the optical frequency of a laser is modulated by a function $a(t)$.
The light reflecting from a moving target returns to the receiver with a delay $\tau$ and Doppler shift $f$ such that the received signal has an instantaneous frequency (IF) $a(t-\tau)-f$~\cite{chang_distance_1997}. 
By interfering the received signal with a local copy of the transmitted signal, the interferometer produces an interference signal with IF
\begin{equation}
\g{t}{\tau}{f} = \begin{cases}
\pm \left[ a(t)-a(t-\tau) + f \right] & \text{real},\\
a(t)-a(t-\tau) + f &\text{complex},
\end{cases}
\end{equation}
where ``complex'' indicates that the receiver measures both in-phase and quadrature (I/Q) channels, whereas ``real'' captures only an in-phase measurement.
When the modulation $a(t)$ is linear (or affine),
the interference signal $\g{t}{\tau}{f}$ contains constant beat frequency (CBF) segments.
In such cases, the estimation of $\tau$ and $f$ can be done by frequency estimation, which is computationally simple and fast. 
Therefore, piecewise linear functions such as sawtooth and triangular waves are the predominant choices for $a(t)$.
In \cref{fig:firstfig}, we show a sketch of the instantaneous frequency for FMCW lidar using triangular modulation. 
The transmitted signal has center frequency $\fc$, and the frequency varies across bandwidth $B$ over the chirp duration $T$.
The instantaneous frequency of the interference signal $\g{t}{\tau}{f}$ contains CBF segments (see \cref{fig:first_b} and \cref{fig:first_c}).

Nyquist~\cite{nyquist_certain_1928} showed---and Shannon~\cite{shannon_communication_1949} formalized---that sampling a band-limited signal at twice the highest frequency is sufficient to avoid aliasing distortion.
While the CBF-based solution to FMCW lidar is elegant, it fails when 
the beat frequency is above 
the Nyquist frequency (half the sampling rate $\fs$), resulting in
aliasing~\cite{kim_longerdistance_by_freqmix_2023}.
Although parameters such as $B$, $T$, or 
$\fs$ can be tuned to avoid this issue, these adjustments come
at the cost of lowering the accuracy of the measurements, lowering the frame rate, or increasing demands on the hardware~\cite{nordin_advantages_2002}.
The CBF-based approach further requires that the modulation be as linear as possible~\cite{lihachev_2022}, so significant effort is often needed to avoid nonlinearities or phase noise, which degrade CBF estimation.
Finally, even if the modulation is linear and no aliasing occurs, CBF-based processing ignores the information in the non-CBF parts of the interference signal.

In this work,
we focus on distance and velocity estimation of a single point reflector 
using
FMCW lidar,
assuming quadrature measurements are available and phase noise is the dominant noise source.
We propose two methods that address limitations of CBF-based
processing by leveraging the entire signal,
avoiding ambiguities due to aliasing
and not depending on linear frequency modulation.
Given a modulation function $a(t)$,
the instantaneous phase and
instantaneous frequency are described by functions parametrized by the delay $\tau$ and Doppler shift $f$.
By leveraging this parametric relationship and accounting for the possibility of aliasing, our proposed methods can recover
a range of distances that are beyond the limit set by the Nyquist frequency and double the range of velocities compared to the conventional method
(see \cref{fig:first_d}). 
Furthermore, since our approach does not require that $\g{t}{\tau}{f}$ contains CBFs, we can choose from a broader range of modulation functions $a(t)$ (e.g., sinusoidal or more arbitrary) and still recover distance and velocity. 
Our first method applies matched filtering directly to the measured interference signal using the parametric instantaneous phase.
Matched filtering is simple but is sensitive to phase noise, especially with triangular modulation.
Our second approach, which we call \emph{instantaneous frequency fitting (IFF)}, first estimates the instantaneous frequency and then fits a parametric function.
Our IFF formulation depends on modeling of the two main sources of noise---shot noise and phase noise---to enable accurate fitting. 
Specifically, we use a wrapped normal approximate likelihood with parameters computed from observable quantities.

We make three main contributions in this paper: 
\begin{enumerate}
\item identifying limitations (e.g., linearity, sampling rate) of the conventional approach to FMCW; 
\item establishing that matched filtering and our proposed  instantaneous frequency fitting method can avoid the distance and velocity limits due to aliasing; and
\item demonstrating that under some conditions, nonlinear modulation can achieve superior performance based on our framework. 
\end{enumerate}

We first review some related work (\cref{sec:related_work}) and the FMCW lidar signal model (\cref{sec:signal_model}), 
followed by a review of the conventional approach to delay and Doppler estimation based on CBFs and its limitations (\cref{sec:cbf_based_processing}). 
We then introduce matched filtering and our 
instantaneous frequency fitting approach
(\cref{sec:fullduration}) and discuss the non-convexity introduced by aliasing (\cref{sec:landscape}). 
We present numerical results obtained from simulated measurements (\cref{sec:num_results}) and compare them with misspecified Cram\'er--Rao bounds for multiple modulation functions (\cref{sec:analysis}).
Finally, we discuss some limitations and potential solutions  (\cref{sec:conclusion}).

\section{Related Work}
\label{sec:related_work}

We begin by detailing some of the challenges and resulting limitations of FMCW lidar.
\cref{tab:fmcw_techniques} presents an overview of the capabilities of the FMCW lidar processing approaches most relevant to our proposed method.
While some specialized hardware systems have been proposed to address individual challenges such as aliasing~\cite{kim_longerdistance_by_freqmix_2023} or arbitrary modulation functions~\cite{pang_arbitrary_2025}, we consider only direct comparisons with systems using a standard FMCW acquisition setup.
Additional degradations such as
speckle,
optical path length drift, and material dispersion effects~\cite{pan_major_2024} are not considered here.

\begin{table}[t]
\caption{Capabilities of FMCW processing techniques. 
A checkmark (\checkmark) indicates good performance, whereas a triangle ($\triangle$) indicates mixed results.}
\label{tab:fmcw_techniques}
\centering
\begin{tabular}{@{}lccccc@{}}
\toprule
\multirow{2}{*}{Method} & \multicolumn{3}{c}{Modulation Function}                   & \multicolumn{2}{c}{Degradation} \\ \cmidrule(l){2-4} \cmidrule(l){5-6} 
                        & Triang.  & Sinus.  & Arb. & Aliasing      & Phase noise     \\ \midrule
Periodogram~\cite{rife_single_1974}             & \checkmark             &                       &           &               &                 \\
Lorentzian~\cite{kim_optimal_2018}       & \checkmark             &                       &           &               & \checkmark       \\
Tsuchida~\cite{tsuchida_freqavg_2019}                &                       & \checkmark             &           &               &                 \\
Matched filter~\cite{turin_introduction_1960}          & \checkmark             & \checkmark             &   \checkmark          & \checkmark     & $\triangle$ \\
IFF (proposed)                & \checkmark            & \checkmark             & \checkmark            & \checkmark     & \checkmark       \\ \bottomrule
\end{tabular}
\end{table}

\subsection{Nonlinearity Correction}\label{sec:nonlinear_correction}
A linear sweep of frequency over time results in CBF regions, from which distance and/or velocity can be easily estimated through simple frequency estimation~\cite{rife_single_1974}.
However, linear modulation of the laser drive current or voltage typically results in a nonlinear frequency sweep.
Numerous approaches exist to improve the hardware linearity~\cite{Riemensberger2020, lihachev_2022},
pre-distort the drive current to achieve frequency-sweep linearity~\cite{zhang_laser_2019, zhang_demonstration_2022},
or calibrate the frequency sweep with time so the measured signal can be re-sampled as if the sweep were linear~\cite{wang_cubic_2016, hariyama_high-accuracy_2018, okano_swept_2020, sun_highly-time-resolved_2023,feng_nonlinear_2025, gong_fmcw_2025}.
Our proposed method requires knowledge of the modulation function $a(t)$, but $a(t)$ need not be piecewise linear, and no linearization or correction is performed.

\subsection{Phase Noise}

Even for linear modulation,
a limitation of swept-source lasers is the purity of the frequency as a function of time~\cite{strzelecki_investigation_1988}.
Random fluctuations in the phase of a laser cause deviations in the optical frequency~\cite{vasilyev_optoelectronic_2013}, which is often exacerbated by frequency modulation.
This spectral spread of the laser is characterized by its \emph{linewidth}. 
Large linewidth is associated with reduced temporal coherence of the laser, which decreases the maximum distance that can be measured and otherwise degrades estimation in FMCW lidar.
Significant efforts have been made to produce lower-linewidth lasers to extend the useful range that can be measured~\cite{lihachev_2022, tang_hybrid_2022}.
Recent work has shown that carefully modeling the noise in swept-source lasers leads to estimation algorithms that can yield improved estimation of distant targets far beyond the laser coherence range~\cite{kim_optimal_2018, ulvog_phase_2023}.
Although degradations due to phase noise are not the focus of the present paper, we include realistic phase noise in our simulations of FMCW measurements.

\subsection{Nonlinear Frequency Modulation}
While
linear functions are the most common frequency modulation for FMCW radar and lidar, nonlinear modulation may have desirable properties.
Some approaches for lidar have used sinusoidal modulation, which may be easier to achieve with laser driver circuitry than linear modulation~\cite{chang_distance_1997}.
For instance, Tsuchida~\cite{tsuchida_freqavg_2019} introduced an elegant approach to estimating distance and velocity from sinusoidal modulation; however,
the simple signal processing requires the delay $\tau$ to be much less than the modulation period $T$
(see \cref{fig:dest}).

In the context of pulsed radar, nonlinear modulation coupled with matched filtering has been shown to generate sharper peaks and be less susceptible to errors~\cite{milleit_milleitwaveform_1970, galati_waveformdesign_2014, galati_hybridnonlinearFM_2015}. 
Matched filtering has also been used for FMCW radar for long-range detection~\cite{cooper_2Tprocess_2017}. 
While matched filtering is capable of delay and Doppler estimation for arbitrary frequency modulation, it has not been intended for the case when the signal is aliased.
Depending on the modulation function, matched filtering can be susceptible to large errors when the signal is aliased due to non-convexity, which we demonstrate and discuss in \cref{sec:num_results}.

Alternatively, Pang et al.~\cite{pang_arbitrary_2025} recently proposed an approach to FMCW lidar that allows for arbitrary (non-linear) waveforms.
Their method uses an auxiliary interferometer to track the instantaneous phase at a fixed path length, in addition to tracking the phase for the unknown distance, and an estimator is constructed based on the spectra of the phase signals.
However, this approach is limited by a low-bandwidth, short-range approximation and assumes zero Doppler velocity, unlike our methods.
Our approach also avoids using an auxiliary interferometer by assuming the frequency modulation $a(t)$ is known.

\subsection{Maximum Range of Distance and Velocity}
The range of
distance and velocity values that can be unambiguously resolved by an FMCW ranging system is limited to what we call the \emph{unambiguous space}.
Various sources of ambiguity have been considered in the literature on FMCW radar and lidar.
For instance, Kim et al.~\cite{kim_longerdistance_by_freqmix_2023} note that ranging precision can be increased by increasing the modulation bandwidth;
however, the sampling rate of the analog-to-digital converter (ADC) would then also need to be increased to avoid distance ambiguity due to aliasing if CBF-based estimation is used.
They instead propose a two-step mixing process, where the second stage mixes the interference signal again with the local oscillator, encoding the delay in a frequency zero-crossing. 
However, their method requires additional mixing hardware and linear frequency modulation.
Another approach by Gai et al.~\cite{gai_non-reconstruction_2025} is to use multiple mutually-delayed ADCs with sub-Nyquist sampling rates to achieve higher resolution from only a subset of the high-rate samples, taking advantage of the sparsity of the CBF signal in the frequency domain.
Alternatively, Wang et al.~\cite{wang_breaking_2026} recently proposed co-prime k-clock resampling and reconstruction based on the Chinese remainder theorem for the case when the maximum distance is limited by the path length of an auxiliary interferometer.
Additional sources of ambiguity include the periodicity of the frequency modulation signal~\cite{liu_ambiguityfmcw_2019}, or if the Doppler shift in real measurements contributes more
to the beat frequency than the delay,
which can be avoided with more complicated modulation schemes~\cite{nordin_advantages_2002, feneyrou_frequency-modulated_2017-1, feneyrou_frequency-modulated_2017-2, banzhaf_phase-coded_2021, na_optical_2023} or I/Q coherent detectors~\cite{xu_fmcw_2019}.

Like Kim et al.~\cite{kim_longerdistance_by_freqmix_2023} and Gai et al.~\cite{gai_non-reconstruction_2025}, we consider the effect of the sampling rate (i.e., aliasing) on the unambiguous space but for existing FMCW lidar architectures.
Our techniques require neither additional hardware, additional measurements, nor strong spatial or temporal priors.
Instead, the unambiguous range is extended by taking advantage of information contained in the signal that is conventionally unused.
We analyze in \cref{sec:unam} the unambiguous space of range and velocity estimation for both real and complex measurements,
and we propose in \cref{sec:fullduration} a solution based on complex-valued measurements.

\subsection{FMCW Lidar with Quadrature Measurements}
FMCW lidar implementations often measure only the in-phase component of the interference signal. 
It is also possible to measure the quadrature component to construct a complex signal.
While the detectors required to measure the quadrature component have more complicated designs, a complex (I/Q) signal offers easier extraction of the phase and also extends the unambiguous space of delay and Doppler~\cite{xu_fmcw_2019}.
The most common approach is to use a $90\degree$ optical hybrid~\cite{seimetz_options_2006, khachaturian_iq_2021}, which splits the local oscillator in two copies,
applies an optical Hilbert transform ($\Frac{\pi}{2}$ phase shift) to one copy, mixes the received signal with each copy, and detects the interference signal
with two pairs of balanced photodiodes. 
Throughout this work, we will assume such a receiver is available.

\section{FMCW Signal Model}
\label{sec:signal_model}

We next review the measurement model for FMCW lidar.
Recall that the modulation function $a(t)$ in the system design represents the desired instantaneous frequency.
Thus, an FMCW ranging system transmits a frequency-modulated wave 
\begin{equation}
\xTX(t) = \sqrt{2\PTX}\cos\!\left[2\pi \int_{0}^{t}a(s) \dif s + b + \omega(t)\right],
\end{equation}
where $\PTX$ is the power of the signal,
$b$ is an arbitrary constant phase offset, and
$\omega(t)$ is the phase noise from the laser.
The received signal $\xRX(t)$ is time delayed by $\tau$ and Doppler shifted by $f$:
\begin{align}
\xRX(t) = \sqrt{2\PRX}\cos \biggl[ 2\pi \int_{0}^{t-\tau} & a(s) \dif s + b \nonumber \\
&+ 2\pi ft+\omega(t-\tau)\biggr].
\end{align}
The delay $\tau$ and Doppler shift $f$ are directly proportional to our parameters of interest, distance $d$ and velocity $v$: 
\begin{equation}
d = \frac{\tau c}{2},
\qquad
v = \frac{fc}{2 \fc},
\end{equation}
where $c$ is the speed of light.
In practice, $\xTX(t)$ can be generated either by directly modulating the frequency of a tunable laser~\cite{vasilyev_optoelectronic_2013, kim_thesis_2020}
or by externally modulating the frequency of a fixed-frequency laser, e.g., using an electro-optic modulator (EOM)~\cite{gao_fmcw_heterodyne_iq_2012,yi_ssb_2021}.

\begin{figure}[t]
\centering
\includegraphics[width=\linewidth]{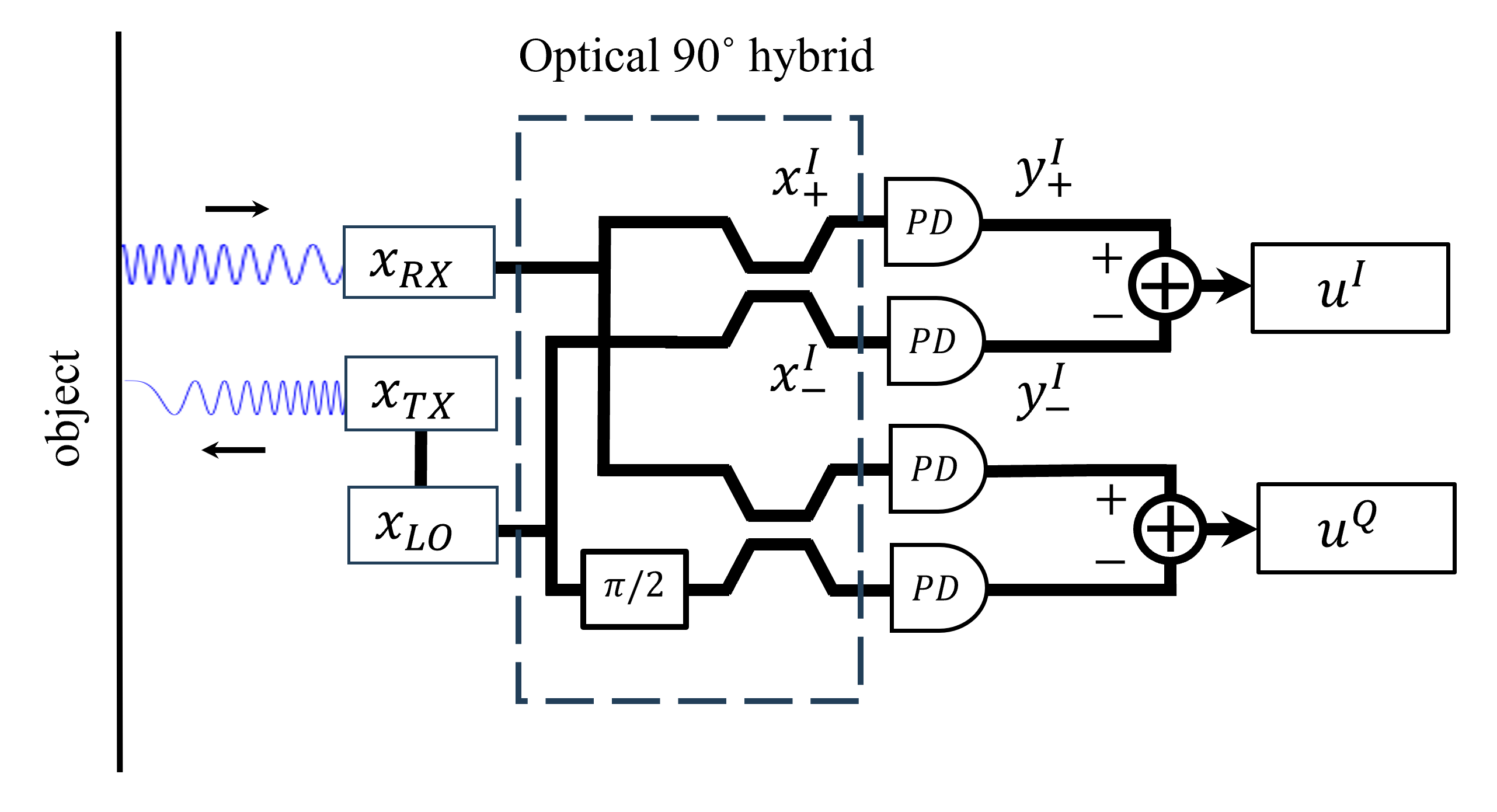}
\caption{
I/Q measurement scheme for FMCW lidar.
The wave $\xTX$ is transmitted toward and reflected from an object.
The received wave $\xRX$ is combined with $\xLO$, a copy of $\xTX$, through an optical $90 \degree$ hybrid consisting of two {\TwoByTwo} couplers and a
$\pi/2$ phase shift. 
Balanced photodetectors (PD) measure the differences between pairs of the coupler outputs, removing the DC components and resulting in an in-phase signal $\uI$ and a quadrature signal $\uQ$.}
\label{fig:schematic}
\end{figure}

The signal flow for I/Q measurements is shown in \cref{fig:schematic}.
Processing of the received signal $\xRX(t)$ uses
local oscillator $\xLO(t)$, which is a copy of $\xTX(t)$.
For the in-phase measurement,
the two signals
are combined through an optical coupler to produce $x^{\rm I}_{+}(t)$ and $x^{\rm I}_{-}(t)$:
\begin{equation}
\begin{bmatrix}
x^{\rm I}_{+}(t) \\ x^{\rm I}_{-}(t)
\end{bmatrix}
= \begin{bmatrix}
   1/\sqrt{2} & \hfill 1/\sqrt{2} \\
   1/\sqrt{2} &       -1/\sqrt{2}
\end{bmatrix}
\begin{bmatrix}
\xLO(t) \\ \xRX(t)
\end{bmatrix}.
\label{eq:coupler}
\end{equation}
Physically,
$x^{\rm I}_{+}(t)$ and $x^{\rm I}_{-}(t)$ are electric fields incident on square-law photodetectors,
which measure intensity while suppressing high frequency terms.
The outputs of the two photodetectors are
\begin{subequations}
\label{eqn:y+y-}
\begin{align}
y^{\rm I}_{+}(t) &= |x^{\rm I}_{+}(t)|^2 \approx \frac{A_2}{2} + \frac{A_1}{2}\cos[\phi(t) + \xi(t)] + \eta^{\rm I}_{+}(t), \\
y^{\rm I}_{-}(t) &= |x^{\rm I}_{-}(t)|^2 \approx \frac{A_2}{2} - \frac{A_1}{2}\cos[\phi(t) + \xi(t)] + \eta^{\rm I}_{-}(t),
\end{align}
\end{subequations}
where 
\begin{align}
 A_1 = 2 \RPD &\sqrt{\PLO \PRX},
\qquad
 A_2 = \RPD(\PLO+\PRX), \\
\phi(t) &= 2\pi \int_{t-\tau}^{t} a(s) \dif s + 2\pi f t, \label{eq:def_phi} \\
\xi(t) &= \omega(t) - \omega(t-\tau),
\end{align}
$\RPD$ is the photodetectors' responsivity,
and $\eta_+(t)$ and $\eta_-(t)$ are shot noise from each photodetector. 
The constant terms in~\eqref{eqn:y+y-} can be canceled by a simple subtraction:
\begin{align}
\uI(t) &= y^{\rm I}_{+}(t) - y^{\rm I}_{-}(t) \nonumber \\
&= A_1 \cos[\phi(t) + \xi(t)] + \etaI(t) ,
\end{align}
where $\etaI(t) = \eta_+(t) + \eta_-(t)$ is the total in-phase shot noise.
The variance of the shot noise is
\begin{equation}
\label{eq:var_eta_I}
\Var[\etaI(t)] = q A_2 = q \RPD (\PTX + \PRX),
\end{equation}
where $q$ is the charge of an electron.

To produce the quadrature component,
a $90\degree$ optical hybrid applies a $\pi/2$ phase shift
to produce $\xLOQ(t)$, which takes the place of $\xLO(t)$ 
in the {\TwoByTwo} coupler input~\eqref{eq:coupler}.
After following the steps above, we get the quadrature component
\begin{equation}
\uQ(t) = A_1 \sin[\phi(t) + \xi(t)] + \etaQ(t) ,
\end{equation}
where the quadrature phase noise variance is also given by \eqref{eq:var_eta_I}.
Combining the in-phase and quadrature components gives the complex signal
\begin{equation}
\label{eq:u(t)-definition}
u(t) = \uI(t) + j \uQ(t)  = A_1 e^{j[\phi(t)+\xi(t)]} + \eta(t).
\end{equation}

A typical assumption is that the laser has frequency noise that is a white Gaussian process~\cite{vasilyev_optoelectronic_2013}. This makes the laser phase noise $\omega(t)$ a Wiener process with variance scaled by the laser linewidth $L$.
As a result, the observed phase noise $\xi(t)$ is a zero-mean stationary Gaussian process that has a triangular autocorrelation function
\begin{equation}
R_{\xi}(u) = \max \left\{ 2\pi L(\tau - |u|), \, 0 \right\}.
\label{eq:phasenoiseautocorr}
\end{equation}
We define a parameter
$\SNReta$
that describes the signal-to-noise ratio with respect to the amplitude (shot) noise:
\begin{equation}
\SNReta = \frac{A_1^2}{\Var(\etaI)+\Var(\etaQ)}
    = \frac{A_1^2}{2 q A_2}
    = \frac{2 \RPD \PLO \PRX}{q (\PLO + \PRX)}.
\end{equation}

\section{CBF-Based FMCW Processing}
\label{sec:cbf_based_processing}

Now that we have introduced the signal model for FMCW lidar, we discuss here the standard CBF-based approach to estimating delay and Doppler and its drawbacks.

\subsection{Linear Frequency Modulation}
Given a modulation function $a(t)$, the IFs of the transmitted and received signal are $a(t)$ and $a(t-\tau)-f$, respectively. 
The IF of the complex-valued interfered signal $u(t)$ is
\begin{equation*}
\g{t}{\tau}{f} = a(t)-a(t-\tau) + f.
\end{equation*}
The most common choices for FMCW ranging are sawtooth modulation and triangular modulation. 
Triangular modulation includes both up and down chirps, and unlike sawtooth modulation, it enables measurement of the  Doppler frequency. 
The modulation function for triangular modulation is defined as
\begin{equation}
a(t) = \sum_k (-1)^k B\left[t-T\left( k + {\textstyle\frac{1}{2}} \right)\right] \mathbbm{1}_{t\in[kT,(k+1)T)]} + \fc,
\label{eqn:trimod}
\end{equation}
where
$B$ is the bandwidth, $T$ is the chirp length, $\fc$ is the center frequency, and $\mathbbm{1}$ is the indicator function.
This triangular frequency modulation is
shown in \cref{fig:first_a}.
Using  
the triangular modulation function $a(t)$, 
the interference consists of four piecewise-linear IF components:
\begin{equation}
\g{t}{\tau}{f}
= \begin{cases}
\hfill 2(\Frac{B}{T})\left(t-\frac{1}{2}\delay\right) + f, & t\in [0, \tau]; \\
\hfill (\Frac{B}{T})\delay + f , & \tau \in [\tau, T]; \\
-2(\Frac{B}{T})\left(t- T -\frac{1}{2}\delay\right) + f, &  \tau \in [T, T+\tau]; \\
\hfill -(\Frac{B}{T})\delay + f , & \tau \in [T+\tau,2T],
\end{cases}
\label{eqn:trimod_freqs}
\end{equation}
which we illustrate 
in \cref{fig:first_c}.
The first and third components are conventionally considered ``error intervals,'' during which no useful measurements can be made~\cite{nordin_advantages_2002},
whereas the second and fourth components are the CBFs 
\begin{equation}
\fbA = \frac{B}{T}\delay + f
\quad
\mbox{and}
\quad
\fbB = -\frac{B}{T}\delay + f.
\label{eq:CBFs-complex}
\end{equation}
Solving for delay $\tau$ and Doppler frequency $f$ requires only simple sums and differences of $\fbA$ and $\fbB$:
\begin{equation}
\label{eq:tau-f-recovery}
\delay = \frac{1}{2}\frac{T}{B}\left(\fbA-\fbB\right)
\quad
\mbox{and}
\quad
f = \frac{1}{2}\left(\fbA+\fbB\right).
\end{equation}

When only the in-phase component (real signal) is available, both positive and negative frequencies are present at once. 
The IF of the real signal $\uI(t)$ is
\begin{equation*}
\g{t}{\tau}{f} = \pm |a(t)-a(t-\tau) + f|.
\end{equation*}
For real signals, it is assumed that $(B/T)\tau > |f|$, so 
only the non-negative frequencies are needed.
Then the CBFs are 
\begin{equation}
\fbA = \left| \frac{B}{T}\delay + f \right|
\quad
\mbox{and}
\quad
\fbB = \left| - \frac{B}{T}\delay + f \right|,
\label{eq:CBFs-real}
\end{equation}
and the delay and Doppler shift are recovered as
\begin{equation}
\label{eq:tau-f-recovery-real}
\delay = \frac{1}{2}\frac{T}{B}\left(\fbA+\fbB\right)
\quad
\mbox{and}
\quad
f = \frac{1}{2}\left(\fbA-\fbB\right).
\end{equation}

Estimating $\fbA$ and $\fbB$ from CBF signals is usually as simple as applying a discrete Fourier transform (DFT) and finding the peak magnitude.
However, the boundaries of the intervals in~\eqref{eqn:trimod_freqs} are uncertain since $\tau$ is unknown a priori.
It is typically
assumed that $\tau \ll T$,
so
$\fbA$ and $\fbB$ can be estimated 
on the intervals $[0,T)$ and $[T,2T)$, respectively, 
despite also containing the linear chirps in the first and third components of~\eqref{eqn:trimod_freqs}.
Alternatively, the intervals $[T_0,T)$ and $[T+T_0,2T)$ can be used, where $0<\tau<T_0<T$ so that the intervals only contain CBF segments over the expected range of $\tau$~\cite{lin_4-d_2023}. 

Recent work has shown that simple peak detection may not be the most effective estimator in FMCW lidar since spectral broadening occurs due to phase noise~\cite{vasilyev_optoelectronic_2013}, and modeling the phase noise can yield improved estimation far beyond the laser coherence range~\cite{kim_optimal_2018, ulvog_phase_2023}.
Under a white frequency noise assumption,
the power spectral density of the phase noise can be approximated by a Lorentzian function parameterized by the laser linewidth~\cite{vasilyev_optoelectronic_2013, kim_thesis_2020}. 
Strategies such as 
fitting a Lorentzian function to the DFT can outperform frequency estimators that only address additive noise.

\subsection{CBF of a Sampled Signal}
\label{sec:conv_drawbacks}

\begin{figure}
\centering
\includegraphics[width=\linewidth]{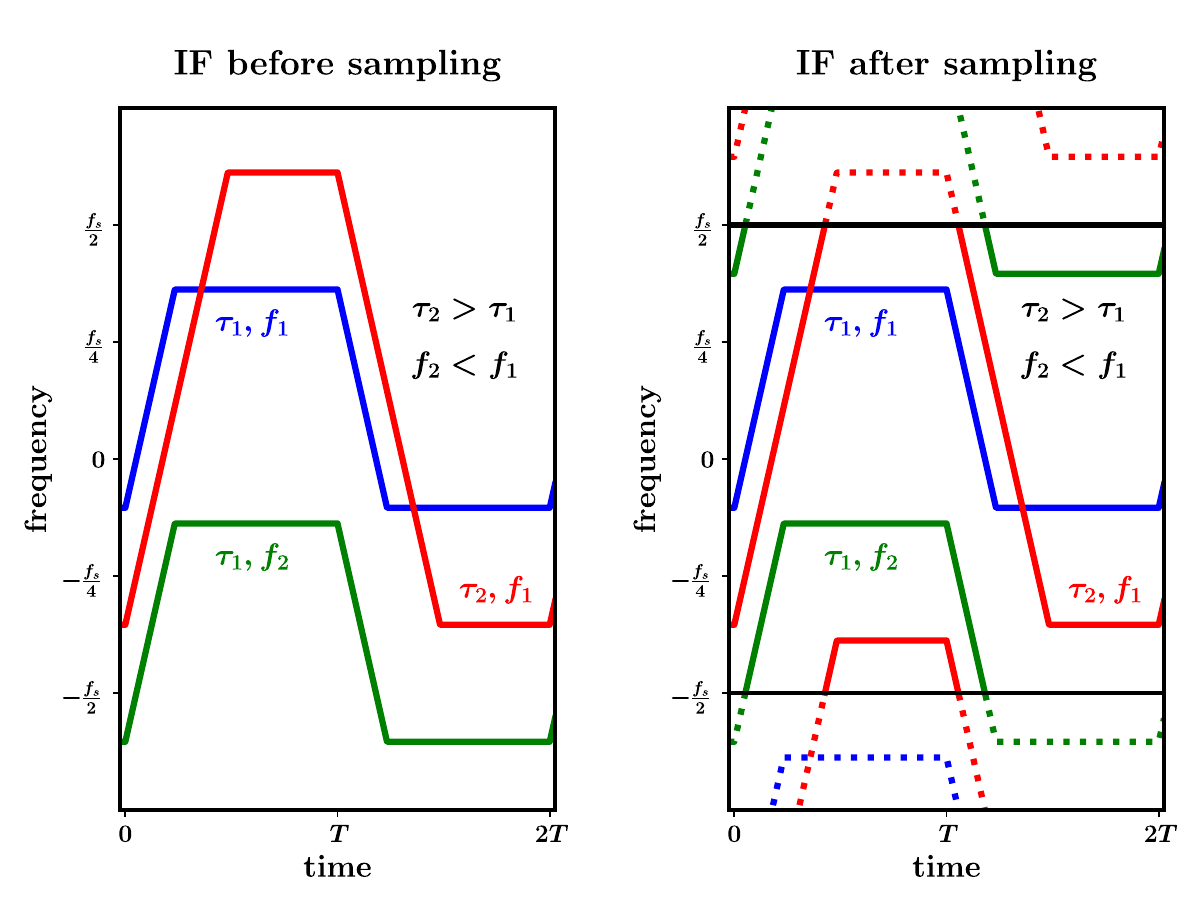}
\caption{The instantaneous frequency (IF) of $u(t)$ for three sets of $(\tau, f)$ before and after sampling. 
Blue and green share the same $\tau$ but different $f$. Blue and red share the same $f$ but different $\tau$. 
Sampling the signal will produce multiple shifted copies of the original IF\@. When measuring the frequencies of a sampled signal, we only observe what is within the Nyquist frequency (from $-\fs/2$ to $\fs/2$). }
\label{fig:if_before_after_sampling}
\end{figure}

Solving for $\delay$ and $\doppler$ is simple for continuous-time signals.
In practice, however, the %
processing is performed on a discrete-time signal
$\vecu = [u(t_0), \dots, u(t_{N-1})]$, where $u(t_n) = u(n/\fs)$.
Due to uniform sampling with rate $\fs$, frequencies beyond the range of $(-\fs /2, \fs/2]$ will be aliased. 
It is more accurate to describe the signal's IF by its ``wrapped'' version, 
\begin{equation}\label{eq:modulo_if}
g^{\text{w}}_{\tau, f}(t) =  \Omega_{\fs}\!\left[ \g{t}{\tau}{f} \right],
\end{equation}
where the centered modulo operator
\begin{equation}
\Omega_r(\cdot) = (\cdot + r/2) \text{ mod } r - r/2
\end{equation}
wraps signals into $(-r/2, r/2]$.
We show $g_{\tau, f}(t)$ and $g^{\text{w}}_{\tau, f}(t)$ for the case of triangular modulation in \cref{fig:if_before_after_sampling}.
The CBFs for complex-valued 
$\vecu$ %
are then
\begin{equation}
\fbAw = \Omega_{\fs}\!\left( \frac{B}{T}\tau+f\right),
\quad
\fbBw = \Omega_{\fs}\!\left(-\frac{B}{T}\tau+f\right),
\end{equation}
or for real-valued samples are
\begin{equation}
\fbAw = \left|\Omega_{\fs}\!\left( \frac{B}{T}\tau+f\right)\right|,
\quad
\fbBw = \left|\Omega_{\fs}\!\left(-\frac{B}{T}\tau+f\right)\right|.
\end{equation}

\begin{figure}
\centering
\includegraphics[trim={23mm 0mm 29cm 0mm}, clip, width=\linewidth]{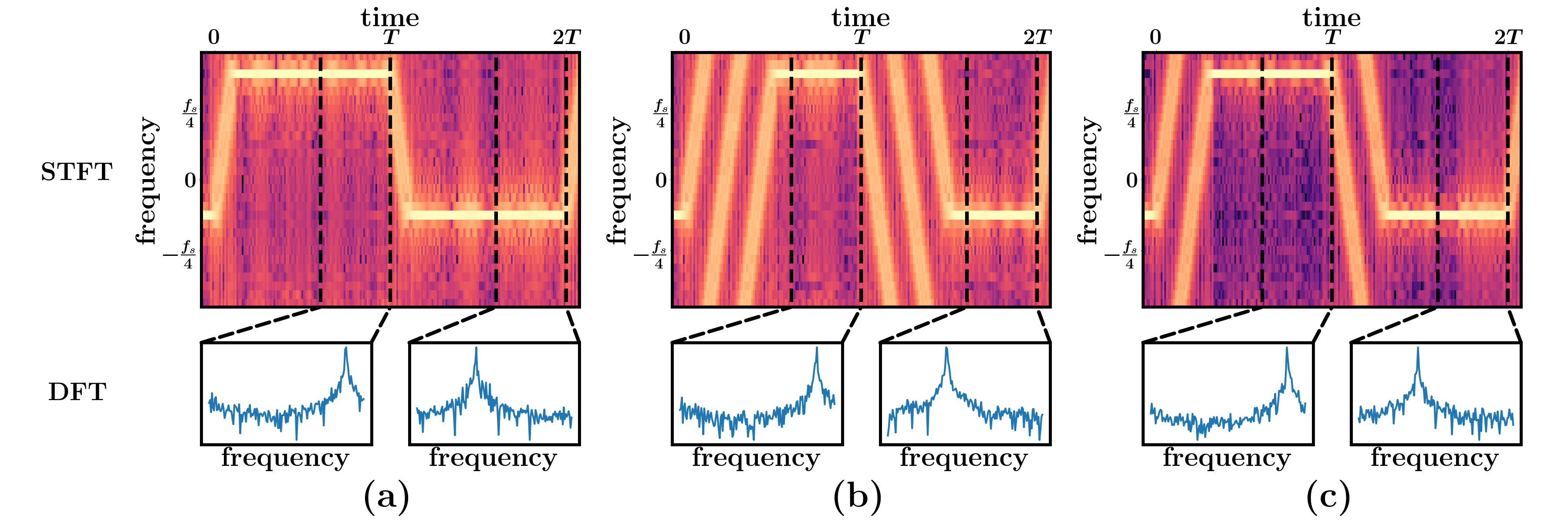}
\caption{Two examples with different $\tau$ and $f$. 
While the examples are indistinguishable from the DFTs over the CBF regions, from the STFTs we clearly see that they are different.}
\label{fig:stft_3cases}
\end{figure}

When aliasing occurs, \eqref{eq:tau-f-recovery} and \eqref{eq:tau-f-recovery-real}
are no longer valid equations to solve for $\delay$ and $\doppler$
because aliasing introduces ambiguity into frequency estimation.
As an example, we show in \cref{fig:stft_3cases} two signals with different pairs of $\delay$ and $\doppler$ that result in the same CBFs. 
Only by looking at the IFs through the short-time Fourier transform (STFT) is it obvious that the signals have been affected by different amounts of aliasing.

\subsection{Unambiguous Parameter Space}
\label{sec:unam}

\begin{figure*}
\centering
{\phantomsubcaption\label{fig:cbf_1a}}
{\phantomsubcaption\label{fig:cbf_1b}}
{\phantomsubcaption\label{fig:cbf_1c}}
\includegraphics[width=\textwidth]{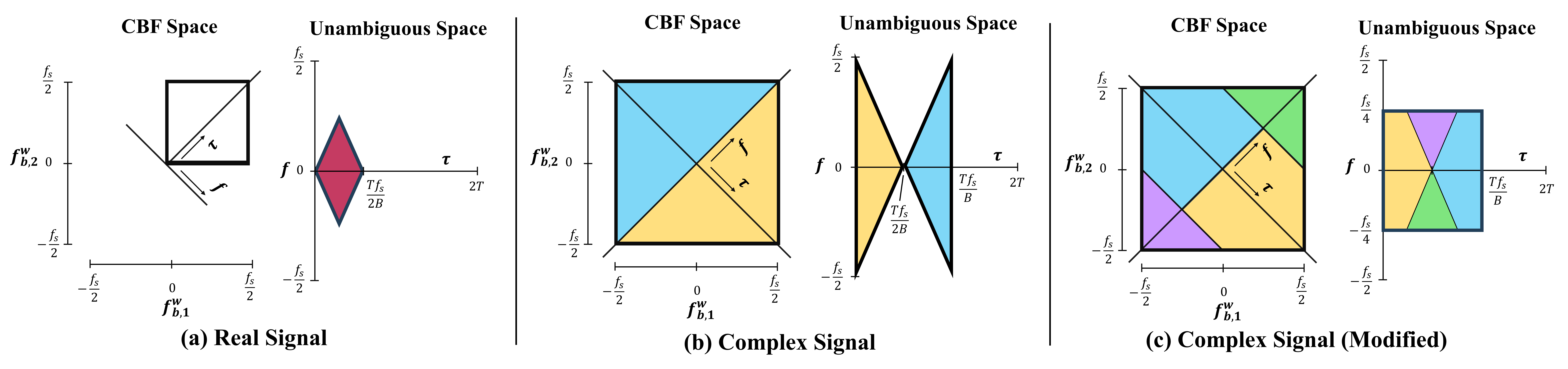}
\caption{The CBF space is the possible set of values for $\fbAw$ and $\fbBw$.
The unambiguous space is the image after mapping from the CBF space.
(a) For real signals, we restrict CBFs to be nonnegative, resulting in an unambiguous space that has a rhombus shape. 
(b) For complex signals, aliasing wraps frequencies in a circle and hence the CBF space is a torus. %
To avoid mapping to negative delay, we could split the CBF space into two triangles, 
resulting in a
a butterfly-shaped unambiguous space. 
(c) In order to resolve the same range of Doppler shifts, regardless of the delay, we propose an alternative partition of the CBF space for complex signals.
The CBF space is divided into four cases as written in \eqref{eq:four_cases}, with corresponding regions shown in the same color. 
A rectangular unambiguous space is more reasonable when non-zero Doppler shift is expected regardless of distance.}
\label{fig:cbf_space1}
\end{figure*}

Since aliasing can lead to ambiguity in the CBF-based approach,
we discuss here the space of $\delay$ and $\doppler$ values that can be recovered \emph{unambiguously}.
We first write the naive conversion of CBFs to $\delay$ and $\doppler$. 
In case of the real signal, based on \eqref{eq:tau-f-recovery-real}, this is
\begin{equation}
\begin{bmatrix}
\tau' \\ f'
\end{bmatrix}
  = \begin{bmatrix}
             \Frac{T}{B} & 0 \\
                       0 & 1
    \end{bmatrix}   
    \begin{bmatrix}
             \Frac{1}{2} & \hfill \Frac{1}{2} \\
             \Frac{1}{2} & -\Frac{1}{2}
\end{bmatrix}   
\begin{bmatrix}
\fbAw \\ \fbBw
\end{bmatrix},
\end{equation}
which is a $45\degree$ rotation, reflection, and rescaling, visualized in \cref{fig:cbf_1a}. Because of the $2T$ periodicity of the modulation function, $2T$ is the upper limit on the delay $\delay$ that can be unambiguously recovered, irrespective of the sampling rate.
Taking the potential aliasing of $\fbAw$ and $\fbBw$ into account, the unambiguous range of $\delay$ for real signals is
\begin{equation}
\taumax - \taumin = \min\left\{\frac{T\fs}{2B},\, 2T\right\},
\end{equation}
and the maximum unambiguous Doppler shift is $(-\fs/4, \fs/4]$.
However, we notice from \cref{fig:cbf_1a} that the unambiguous Doppler shift depends on the delay $\delay$, in order to avoid the ambiguity from both positive and negative frequencies in real-valued measurements.
The total area of the unambiguous space for real signal case is
\begin{equation}
(\taumax-\taumin)\times(\fmax-\fmin) =\min\left\{\frac{T\fs}{2B},\, 2T\right\}\times \frac{\fs}{4}.
\end{equation}

In the case of complex signals, naive conversion of the CBFs could be performed based on \eqref{eq:tau-f-recovery}. 
However, we notice that a negative $\tau'$ value is infeasible for lidar applications. 
Based on this observation, %
a potential remapping is
\begin{align}
\begin{bmatrix}
\tau' \\ f'
\end{bmatrix} = \begin{cases}
\begin{bmatrix}
\frac{1}{2}\frac{T}{B} & -\frac{1}{2}\frac{T}{B} \\
\frac{1}{2} & \frac{1}{2}
\end{bmatrix}
\begin{bmatrix}
\fbAw \\\fbBw
\end{bmatrix}, 
& \text{if } \fbAw > \fbBw; \\[10pt]
\begin{bmatrix}
\frac{1}{2}\frac{T}{B} & -\frac{1}{2}\frac{T}{B} \\
\frac{1}{2} & \frac{1}{2}
\end{bmatrix}
\begin{bmatrix}
\fbAw + \fs \\\fbBw - \fs
\end{bmatrix}, 
& \text{if } \fbAw \leq \fbBw,
\end{cases}%
\label{eq:operator2}
\end{align}
as visualized in \cref{fig:cbf_1b}.
In the second case, we assumed that values of $\fbCw$ are results of aliasing, since otherwise it is impossible to obtain $\fbAw \leq \fbBw$. %

What makes the complex signal case significantly different from the real signal case is the topology of $\fbAw$ and $\fbBw$. 
We refer to the possible set of values for $\fbAw$ and $\fbBw$ as the \emph{CBF space},
which is the rectangular space $[0, \Frac{f_s}{2})^2$ for real signals. 
For a complex signal, since we can observe negative frequencies, the CBF space is $[-\Frac{f_s}{2}, \Frac{f_s}{2})^2$.
The CBF space for complex signals has the topology of a torus (a Cartesian product of circles \cite{armstrong_1983_basic}), 
a frequency increasing past $f_s/2$ will wrap to
$-f_s/2$.
This property makes handling aliasing easier than in the real-signal case. 

We observe in \cref{fig:cbf_1b} that the remapping in \eqref{eq:operator2} causes a butterfly-shaped unambiguous space, i.e., the resolvable Doppler shift depends on the delay.
Instead, we can take advantage of the topology of the CBF space to first translate $\fbCw$ according to
the following four cases:
\begin{subequations}
\label{eq:four_cases}
\begin{align}
    &\text{if } \begin{Bmatrix}
        \fbAw > \fbBw\\
        \fbAw + \fbBw \leq \Frac{\fs}{2} \\
        \fbAw + \fbBw \geq -\Frac{\fs}{2} \\
    \end{Bmatrix}, &\begin{bmatrix}
\fbAw \\\fbBw
\end{bmatrix} &\rightarrow \begin{bmatrix}
\fbAw \\\fbBw
\end{bmatrix}; \label{eq:yellow} \\
&\text{if } \begin{Bmatrix}
        \fbAw \leq \fbBw\\
        \fbAw + \fbBw \leq \Frac{\fs}{2} \\
        \fbAw + \fbBw \geq -\Frac{\fs}{2} \\
    \end{Bmatrix}, &\begin{bmatrix}
\fbAw \\\fbBw
\end{bmatrix} &\rightarrow \begin{bmatrix}
\fbAw +\fs\\\fbBw - \fs
\end{bmatrix}; \label{eq:blue}\\
&\text{if }~\quad\begin{matrix}
        \fbAw + \fbBw > \Frac{\fs}{2} 
    \end{matrix}, &\begin{bmatrix}
\fbAw \\\fbBw
\end{bmatrix} &\rightarrow \begin{bmatrix}
\fbAw \\\fbBw  -\fs
\end{bmatrix}; \label{eq:green}\\
&\text{if } ~\quad\begin{matrix}
        \fbAw + \fbBw < -\Frac{\fs}{2} 
    \end{matrix}, &\begin{bmatrix}
\fbAw \\\fbBw
\end{bmatrix} &\rightarrow \begin{bmatrix}
\fbAw+\fs \\\fbBw
\end{bmatrix}. \label{eq:purple}
\end{align}
\end{subequations}
After translating the CBF vector, we multiply it by the matrix from \eqref{eq:operator2}. 
We show in the illustration in \cref{fig:cbf_1c} how the above four cases map to $\delay$ and $\doppler$ values, where the color correspondences yellow~\eqref{eq:yellow}, blue~\eqref{eq:blue}, green~\eqref{eq:green}, and purple~\eqref{eq:purple} 
result in a rectangular unambiguous space.
Note that the total area of the unambiguous space,
for the complex signal case 
\begin{equation}
(\taumax-\taumin)\times(\fmax-\fmin) =\min\left\{\frac{T\fs}{B},\, 2T\right\} \times \frac{\fs}{2}
,
\end{equation}
is twice that of the real case and unchanged by the mapping in \eqref{eq:four_cases}. 
By translating $\fbCw$ by $\fs$ differently for various cases, we can construct many ways to convert $\fbCw$ to $\delay$ and $\doppler$ that all result in an unambiguous space with the same area. 
This demonstrates the limitation of CBF-based processing imposed by the sampling rate.

\subsection{Other Drawbacks of CBF-based Processing}

When the range is limited by the sampling rate $\fs$, the only option for expanding the unambiguous space is to lower the chirp rate $\Frac{B}{T}$.
However, this introduces its own set of tradeoffs.
When the DFT is applied to a signal of duration $T$, the resolution of the frequency bin is $1/T$. 
In our context, this results in resolution of $1/B$ for delay $\delay$ and $1/T$ for Doppler $f$.
Increasing $B$ will increase the accuracy of estimating $\delay$ and increasing $T$ will increase the accuracy of estimating $f$.

We can also find this result from an estimation-theoretic perspective. 
By assuming the signal is dominated by CBF segments corrupted by additive white Gaussian noise, 
we show in \cref{apx:crb_cbf} that
the Cram\'er--Rao bounds for unbiased estimates $\hat{\delay}$ and $\hat{\doppler}$ are proportional to
\begin{align*}
\Var(\hat{\tau}) \propto \frac{1}{B^2 T \fs}, \quad\Var(\hat{f}) \propto \frac{1}{T^3 \fs}.
\end{align*}
Since the number of samples is on the order of $T\fs$, the Fisher information per sample about $\tau$ and $f$ are on the order of $B^2$ and $T^2$, respectively.
However, we cannot blindly increase $B$ and $T$ since increasing $T$ results in a lower frame rate and increasing $B$ without also increasing $T$ results in higher chance of aliasing ($|(\Frac{B}{T})\tau\pm f|>\Frac{\fs}{2}$).
It also may not be possible to increase $B$ due to the bandwidth limitations of tunable lasers~\cite{nordin_advantages_2002}. 
Aside from the limited unambiguous space imposed by the sampling rate, CBF-based processing suffers from two additional limitations that we highlight here.

\subsubsection{Sample inefficiency}
As the delay becomes longer, the fraction of the signal that exhibits CBFs becomes shorter. 
Only extracting information from the CBFs ignores the information about $\delay$ and $\doppler$ contained in the non-CBF regions.
While in many applications the delay is assumed to be significantly shorter than the modulation period, $\tau \ll T$,
when high frame rate (lower $T$) is desirable, this assumption may
not hold.

\subsubsection{Dependence on linear frequency modulation}
Conventional processing for FMCW ranging relies on accurate linear FM---otherwise there will be no CBFs to estimate.
However, as discussed in \cref{sec:nonlinear_correction}, precise frequency control is a challenging task, 
adding hardware and/or computational complexity
while still being held back by sample inefficiency and limited range.

\section{FMCW Beyond Nyquist}
\label{sec:fullduration}

To resolve a larger $(\delay,f)$ space without changing the transmission scheme,
we propose two signal processing approaches that can perform FMCW lidar estimation beyond the Nyquist frequency limit.
Our first approach is a straightforward adaptation of matched filtering, which relies on a parametric model of the instantaneous phase. 
Our second approach develops a more sophisticated, gradient-based estimation method that relies on a parametric model of the instantaneous frequency.

Both approaches make use of the full signal
$\vecu$
and assume the modulation function is known.
In addition to
triangular modulation with $a(t)$, as defined in \eqref{eqn:trimod},
we also consider sinusoidal modulation, which has been used in a limited number of cases~\cite{tsuchida_freqavg_2019}.
Given the same bandwidth $B$ and chirp duration $T$, the sinusoidal modulation function is
\begin{equation}
a(t) = -\frac{B}{2}\cos\!\left( \frac{\pi t}{T}\right) + \fc.
\end{equation}
Our approaches also extend to more arbitrary modulation functions.

\subsection{Matched Filtering}
\label{sec:mf_method}
The most basic approach to CBF-based processing is the periodogram maximization approach of Rife and Boorstyn~\cite{rife_single_1974}, which seeks the least-squares estimate of a beat frequency.
Given a measurement vector 
$\vecu$,
the maximum periodogram estimate for beat frequency $\fbC$ with $i\in \{1,2\}$ is
\begin{equation}\label{eq:max_periodogram}
\fbC = \argmax_{f} \left|
\sum_{\substack{n:\\t_n\in[(i-1)T,iT)}} \exp(-j2\pi f t_n) u(t_n)
\right|^2.
\end{equation}
CBF estimates can then be converted to distance and velocity estimates following~\eqref{eq:tau-f-recovery}.
The maximum periodogram approach can be considered a type of matched filtering that projects the measurement onto complex exponential basis functions with different frequencies and chooses the frequency resulting in the largest inner product.
This projection ignores the effect of the non-CBF region of the measurement; moreover, if aliasing occurs, the projection cannot distinguish between aliased and unaliased CBFs.

A natural extension of the maximum periodogram approach is least squares estimation of the delay and Doppler parameters directly, without the intermediate step of estimating CBFs.
Given a modulation function $a(t)$, we derived in~\eqref{eq:def_phi} that the instantaneous phase has a known form $\phi(t;\delay,\doppler)$ that depends on the delay $\delay$ and Doppler shift $\doppler$.
Following the literature on matched filtering, we can project the measurements onto functions with different parameters $(\delay,\doppler)$.
Put another way, the matched filtering estimators are
\begin{equation}
    \hat{\delay}, \hat{\doppler}  = \argmax_{\delay, \doppler} \LMF(\vecu;\tau,f),
\end{equation}
where the objective function is
\begin{equation}\label{eq:matched_filter}
    \LMF(\vecu;\tau,f) =\left| \sum_{n} u(t_n) \exp[-j\phi(t_n;\tau,f)] \right|^2.
\end{equation}
The phase $\phi(t;\tau,f)$ can be written in the form
\begin{align}
    \phi(t;\tau,f) &=  2\pi \left[\int_{0}^t a(s)ds - \int_{0}^{t-\tau} a(s)ds+ft\right] \nonumber \\
    &=\phi_0(t) - \phi_0(t-\tau) + 2\pi ft, \label{eq:phase_decomposition}
\end{align}
where $\phi_0(t)$ is the transmitted phase and $\phi_0(t-\tau)- 2\pi ft$ is the received phase. 
The decomposition of the phase
in \eqref{eq:phase_decomposition}
allows us to write the multiplication in
each term of \eqref{eq:matched_filter}
as
\begin{equation}
    u(t_n) e^{-j\phi(t_n;\tau,f)} = \left[u(t_n)e^{-j\phi_0(t_n)}\right] \left[ e^{j\phi_0(t_n-\tau)+j2\pi ft} \right],
\end{equation}
where the second factor is the ``filter'' that depends on $\delay$ and $\doppler$.

\subsubsection{Matched Filter Implementation}
One advantage of matched filtering is a straightforward implementation.
In general, we use a two-step process.
First, the measurements are projected onto filters selected for a discrete parameter grid.
Second, the discrete parameters leading to the maximum cross-correlation are refined via continuous optimization.
Further implementation details are as follows.

\paragraph{Distance-Only Matched Filtering}
When zero-Doppler shift is assumed ($f=0$), 
evaluating
$\LMF(\vecu; \delay, 0)$
for a candidate delay
$\delay=t_n$
is simply a cross-correlation of two discrete signals, which can be computed via an FFT\@.
We expect to observe a peak centered near the true time delay where the width of the peak is approximately the inverse of the frequency modulation bandwidth $1/B$. 
Since in our case the bandwidth of the signal may be larger than the sampling rate, evaluating $\LMF$ only at $N$ points separated by the sampling period can lead to the peak being missed. 
To avoid such missed detections,
upsampling is essential.
We evaluate $\LMF$ on a finer grid
$\tau=m/(M\fs)$,
 $m=0,\dots,MN-1$, where $M = \lceil\Frac{B}{\fs}\rceil$. 
Evaluation on this finer grid can also be quickly performed using an FFT by first upsampling 
$\vecu$ %
by a factor $M$.
After obtaining the magnitude-maximizing argument $\hat{\tau}_{\text{coarse}}$ over the evaluated grid, we refine the estimate further by Brent's method~\cite{brent_algorithms_1973} with initial point $\hat{\tau}_{\text{coarse}}$ and bounds $(\hat{\tau}_{\text{coarse}}-\frac{1}{M\fs}, \hat{\tau}_{\text{coarse}}+\frac{1}{M\fs})$.

\paragraph{Distance--Velocity Matched Filtering}
Allowing for non-zero Doppler requires evaluating
$\LMF(\vecu; \tau, f)$
over a two-dimensional grid.
We use
Doppler candidates
$f = \ell/(2T)$, $\ell = 0,\ldots,N-1$,
and delay candidates as above,
$\tau = m/(M\fs)$, $m = 0,\ldots,MN-1$, where $M = \lceil\Frac{B}{\fs}\rceil$,
thus evaluating
$\LMF(\vecu; \tau, f)$
at a total of $MN^2$ points.
Similar to the delay-only search, we use the Nelder--Mead algorithm~\cite{nelder_1965} to search a finer space with the initial point being the estimate obtained from the grid search and restricting the search space within the adjacent grid points.

\subsubsection{Matched Filtering Limitations}
In pulsed Doppler radar, the large wavelength leads to low sensitivity to the Doppler shift.
The Doppler shift can thus be neglected when estimating $\tau$ by matched filtering, and the velocity can instead be estimated subsequently by tracking a change in phase over multiple chirp  periods \cite{richards_frs3_2022}. 
However, due to the much shorter wavelength in lidar, the Doppler shift is much more sensitive to velocity
and cannot be neglected in matched filtering.
Joint distance--velocity estimation thus requires a two-dimensional search over both $\delay$ and $\doppler$ parameters, resulting in $N$ times more operations than in a delay-only search.

Another limitation of matched filtering is that it is intended for noise added to the signal amplitude; however, phase noise is often more significant in FMCW lidar.
We show through numerical experiments in \cref{sec:num_results} that the resulting random phase variations with time can degrade the matched filtering performance.

\subsection{Instantaneous Frequency Fitting}

Our second approach avoids some of the limitations of matched filtering by instead operating on the instantaneous frequency.
From~\eqref{eq:modulo_if}, the wrapped IF given $(\delay, \doppler)$ and subject to the sampling rate $\fs$ is
\begin{equation}\label{eq:parametric_wrapped_if}
\gwtf(t) = \Omega_{\fs} \! \left[a(t)-a(t-\tau) + f \right].
\end{equation}
Because of the periodicity of the frequency modulation and aliasing caused by the sampling rate,
$g_{\tau,f}^\text{w}(t)$ is periodic with respect to all arguments:
\begin{subequations}
\label{eq:periodicities}    
\begin{align}
g_{\tau,f}^\text{w}(t) &= g_{\tau,f}^\text{w}(t+2T), \label{eq:periodicity_t} \\
g_{\tau,f}^\text{w}(t) &= g_{\tau+2T,f}^\text{w}(t), \label{eq:peridocity_tau} \\
g_{\tau,f}^\text{w}(t) &= g_{\tau,f+\fs}^\text{w}(t). \label{eq:periodicity_f}
\end{align}
\end{subequations}
If we are able to associate $g_{\tau,f}^\text{w}(t)$ evaluated at a sequence of points $t=t_n'$ (not necessarily $t_n$)
to a unique $(\delay, f)$ pair, there is a potential to increase our unambiguous space to a much larger set $(0, 2T] \times (-\fs/2, \fs/2]$
as shown in \cref{fig:first_d}, making the range of solvable distances independent of $B$ and $\fs$.

By characterizing the noise affecting a measurement $u(t)$ and assuming $a(t)$ is known, we would ideally derive a maximum likelihood estimator
\begin{equation}\label{eq:aml}
\hat{\delay}, \hat{\doppler} = \argmax_{\delay, \doppler} \mathcal{L}(\vecu; \delay, \doppler)
\end{equation}
that takes advantage of the parametric model~\eqref{eq:parametric_wrapped_if}.
However, we note several challenges with this approach.
First, the parametric model is in terms of the IF, which needs to be extracted from the measurements.
Second, precise probabilistic modeling of the noise is non-trivial due to multiple sources (shot and phase noise) and temporal correlations in the phase noise.
Finally, aliasing results in a highly non-convex optimization problem, which could result in sub-optimal estimates or long computation times.
In this section, we describe the assumptions we use to derive an approximate noise model and approximate maximum likelihood estimator.
We also describe how to efficiently solve the resulting optimization problem.

\subsubsection{Instantaneous Frequency Extraction}

Recall from \eqref{eq:u(t)-definition} that 
$\vecu$ %
is a signal generated by the interferometer, sampled at frequency $\fs$.
Assuming a single echo in our observed signal, we can estimate the IF from the sampled signal through phase differentiation~\cite{Boashash:92b} as
\begin{equation}
\zeta_n = \frac{1}{2\pi}\text{arg}\left\{ u(t_{n+1})\bar{u}(t_n)\right\},
\quad n = 0,1,\ldots,N-2,
\label{eqn:zeta}
\end{equation}
where $\bar{u}$ denotes the complex conjugate.
The forward finite difference approximation to the derivative causes a half-sample group delay.
Let $\tilde{g}^\text{w}_{\tau,f}(t) = (1/\fs) g^\text{w}_{\tau,f}(t)$ be a normalized IF function.
We aim to solve for $\tau$ and $f$ by minimizing a distance between the IF estimate $\zeta_n$ and the noiseless model-based IF prediction $\tilde{g}_{\tau,f}^\text{w}(t_n')$ with $t_n' = \frac{1}{2}(t_n+t_{n+1})$ in between the measured samples.

We assume $\zeta_n$ can be decomposed as the modulo of the sum of the normalized IF function and independent noise terms 
\begin{equation}
\label{eq:zeta_n_decomposition}
\zeta_n
  \approx \Omega_1\!\left[\tilde{g}_{\tau,f}^\text{w}(t'_n)+ \chi_n +\epsilon_n\right],
\end{equation}
where
\begin{equation}
    \chi_n = \frac{1}{2\pi}(\xi_{n+1}-\xi_{n}) \label{eq:chi_n}
\end{equation}
is transformed from phase noise and
\begin{equation}
    \epsilon_n = \frac{1}{2\pi}\Omega_{2\pi}[\angle(A_1+\eta_{n+1}) - \angle(A_1+\eta_n)] \label{eq:projected_amplitude_noise}
\end{equation}
is transformed from amplitude noise, with $A_1$ the amplitude of the original signal defined in \eqref{eqn:y+y-}.

\subsubsection{Approximate Maximum Likelihood}
The precise negative log-likelihood of the IF sequence $\bm\zeta = [\zeta_0, \dots, \zeta_{N-1}]$ is not available in closed form due to the combination of shot noise and phase noise.
We instead fit a parametric function to the extracted  instantaneous frequency using 
a finite-sum approximation based on a wrapped normal distribution: 
\begin{align}\label{eq:LWN}
    &\LIFF(\bm\zeta;\tau,f)\nonumber\\
    &\quad\triangleq  \sum_{n=0}^{N-2} \log \sum_{k=-K}^{K} \exp\!\left\{ -\frac{[\zeta_n- \tilde{g}^\text{w}_{\tau,f}(t_n')-k]^2}{2{\sigma}^2}\right\},
\end{align}
where $\sigma^2$ is a parameter that describes the noise variance and $K$ is a truncation level chosen for the approximation.
Though increasing $K$ improves the approximation quality,
as a rule of thumb we posit that $K = \lceil 3\sigma \rceil$ is large enough.
This is based on including a number of wrappings that covers at least 99.7\% of the unwrapped noise distribution.
We remark that
\eqref{eq:LWN} implicitly treats each noise sequence $\bm\chi$ and $\bm\epsilon$ as independent and identically distributed (i.i.d.). 
However, they are correlated across time 
through the first-order difference equations \eqref{eq:chi_n} and \eqref{eq:projected_amplitude_noise} and the correlation of the phase noise $\bm\xi$. 
We discuss the correlation of the noise terms in more detail in \cref{sec:cov_mat_approx}.

\subsubsection{Variance Approximation}
\label{sec:variance_approx}

The parameter $\sigma^2$ represents the noise variance in the absence of wrapping.
We can approximate $\sigma^2 \approx \Var(\chi_n + \epsilon_n)$ when the right hand side is reasonably smaller than the modulus range as we will discuss below.
Specifically, we use the approximation
\begin{align}
\widehat{\sigma}^2 = \frac{1}{\pi\fs}L + \hat{h}(\SNRetaHat),
\label{eqn:sigmahat}
\end{align}
where $\hat{h}$ approximates the variance contribution due to additive noise.
We detail our derivations and approximations as follows.

\paragraph{Phase Noise Contribution}
We assume the delay $\tau$ is larger than the time between samples $\Frac{1}{\fs}$.
Given the phase noise autocorrelation~\eqref{eq:phasenoiseautocorr} and the first-differencing operation~\eqref{eq:chi_n}, the variance in the IF estimate due to phase noise is
\begin{align}
        &\Var(\chi_n) = \left(\frac{1}{2\pi}\right)^2 \Var(\xi_{n+1}-\xi_n) \nonumber \\
        & \hspace{1em} = \frac{1}{4\pi^2} \left[\Var(\xi_{n+1}) + \Var(\xi_{n}) + 2\Cov(\xi_{n+1}, \xi_{n})\right] \nonumber \\
        & \hspace{1em} = \frac{1}{4\pi^2}[2\pi L \tau + 2\pi L \tau - 4\pi L (\tau - \Frac{1}{\fs})] \nonumber \\
        & \hspace{1em} = \frac{L}{\pi \fs}.
\end{align}
Conveniently, the first difference operation removes the dependence of the variance on the delay $\tau$, unlike when operating on the instantaneous phase.

\paragraph{Additive Noise Contribution}
The effect of additive noise on the IF estimate is less straightforward.
If $\bm\eta$ is approximately Gaussian, then $\angle(A_1+\bm\eta)$---sometimes referred to as additive observation phase noise~\cite{fu_phase-based_2013, ulvog_phase_2023}---has a projected normal distribution. 
When $\Frac{A_1^2}{\eta^2} \gg 1$, then from Tretter~\cite{tretter_estimating_1985},
$$
\Var\left[\angle(A_1+\eta_n)\right] \approx \frac{\Var(\eta_n)}{2A_1^2}.
$$
Therefore, the variance of $\epsilon_n$ at high SNR can be approximated as
\begin{equation}
\Var(\epsilon_n) \approx \frac{\Var(\eta_n)}{(2\pi)^2A_1^2} = \frac{1}{(2\pi)^2\SNReta}.
\label{eqn:varep_approx}
\end{equation}

However, $\SNReta$ is not observable from $u(t_n)$ alone. 
We assume we can obtain an auxiliary signal $v(t_n)$, formulated similarly to $u(t_n)$ but defined as the sum of photodetector outputs instead of the difference:
\begin{align}
v(t_n) &= \left[\yI_{+}(t_n) + \yI_{-}(t_n)\right] + j \left[\yQ_{+}(t_n) + \yQ_{-}(t_n)\right] \nonumber\\
&= A_2 + \eta'(t_n),
\end{align}
where $\eta'(t) = \eta_+(t) + \eta_-(t)$.
Thus $v(t_n)$ is a constant signal corrupted by the shot noise $\eta'(t)$,
which has the same variance as $\eta(t)$.
Using both $\bm u$ and $\bm v$, we can estimate $\SNReta$ with
\begin{equation}
\SNRetaHat = \frac{|| \bm u||_2^2-\Var( \bm v)}{\Var( \bm v)}.
\label{eqn:snrhat}
\end{equation}

\begin{figure}[t]
\centering
\includegraphics[width=\linewidth]{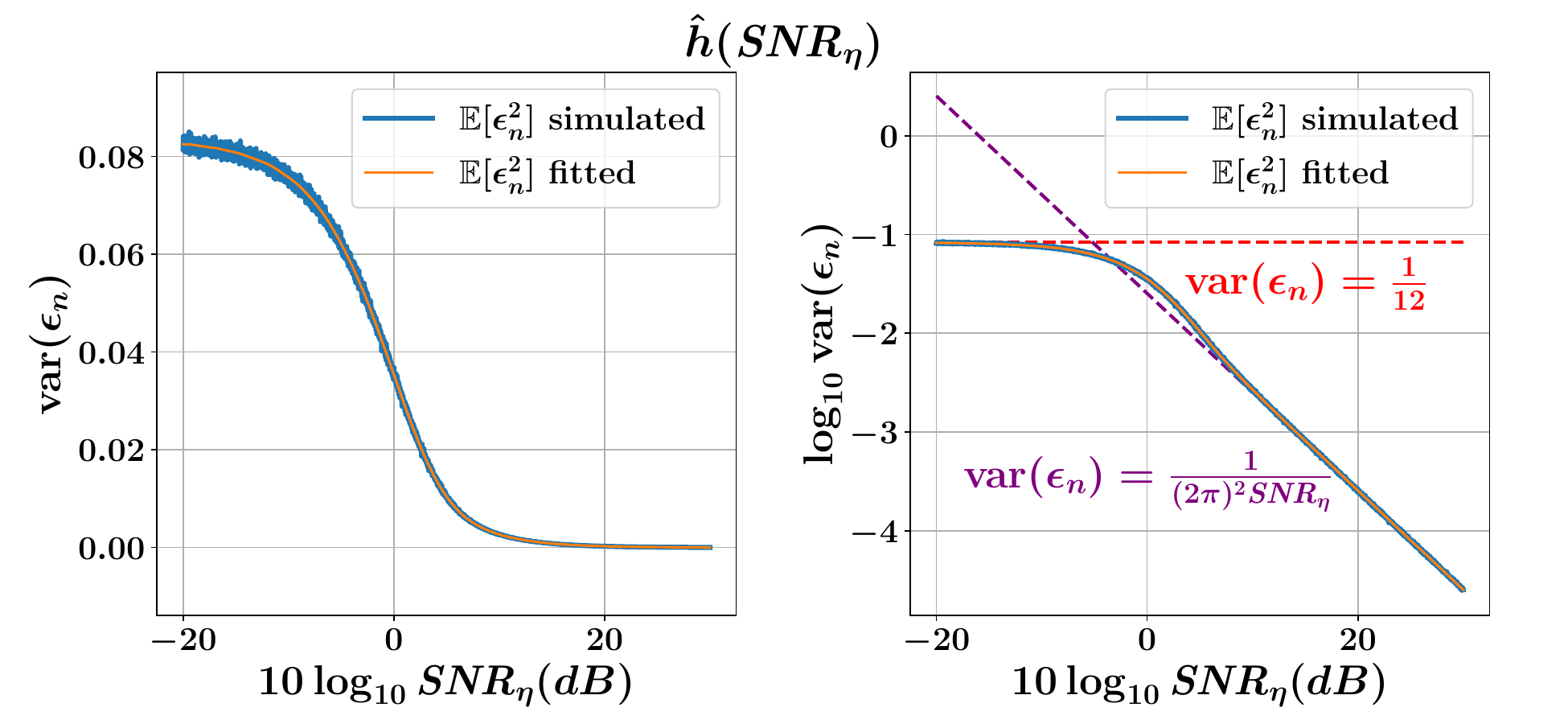}
\caption{
We simulate additive noise $\eta_n$ for varying $\SNReta$, which is then used to generate $\epsilon_n$ and compute $\Var(\epsilon_n)$. 
We show two plots, with and without a log scale in the $y$-axis. The right asymptote is derived in~\eqref{eqn:varep_approx}, and the left asymptote is the variance of the uniform random variable $\Frac{1}{12}$. 
A function $\hat{h}$ fitting smoothing B-splines to the noisy Monte Carlo simulation results is used to compute $\sigma^2$ for $\LIFF$.}
\label{fig:hhat}
\end{figure}

For more accurate estimation of $\Var(\epsilon_n)$, we rely on Monte Carlo simulation of projected normal random variables to regress $\Var(\epsilon_n)$ over $\SNReta$. 
For each value of $\SNReta$, we simulate an $\bm\epsilon$ sequence of length 10\,000 and compute its variance.
Specifically, following \eqref{eq:projected_amplitude_noise},
we generate
\begin{align*}
    \epsilon_n &= \frac{1}{2\pi}\Omega_{2\pi}\left[ \angle \left(1 + \sqrt{\frac{1}{2\SNReta}}z_{n+1}\right) \right.\\
    & \qquad\qquad\qquad\qquad\left. - \angle \left(1 + \sqrt{\frac{1}{2\SNReta}}z_n\right) \right],
\end{align*}
where $z_n$ is an i.i.d.\ complex Gaussian noise sequence.
The final form of 
$\sigma^2$ is
the variance estimate is $\Var(\epsilon_n) \approx \hat{h}(\SNRetaHat)$, where
$\hat{h}$ is shown in \cref{fig:hhat}. 
We see that for $\SNReta$ larger than $\SI{10}{\decibel}$, the curve tends to zero approximately as~\eqref{eqn:varep_approx}.
The left asymptote ($\SNReta < \SI{-10}{\decibel}$) is the variance of a uniform distribution over the phase domain.
For intermediate SNR values (e.g., between \SI{-10}{\decibel} and \SI{10}{\decibel}), a curve fitted to the Monte Carlo simulation results can provide a more accurate estimate of $\hat{h}$.

\section{Optimization Strategy}
\label{sec:landscape}

The maximization of the log-likelihood $\LIFF$ is complicated by
multimodality due to the sum of exponentials
as well as the periodic dependence on $\delay$, $f$, and $t_n$.
Here we study $\LIFF$ to establish how to use gradient-based optimization algorithm from well-spaced initial points to perform the maximization.

\subsection{Contours of Deterministic Variant of $\LIFF$}

$\LIFF$ depends on the IF estimate sequence of the measurement, which is stochastic. Here we consider a deterministic variant of $\LIFF$ that measures the distance between two wrapped IF sequences parameterized by $(\tau_1, f_1)$ and $(\tau_2, f_2)$, respectively:
\begin{equation}
\begin{split}
    &\mathcal{D}(\tau_1,f_1, \tau_2,f_2) = \sum_{n=0}^{N-2} \Omega_1\!\left( \tilde{g}^\text{w}_{\tau_1, f_1}(t_n') - \tilde{g}^\text{w}_{\tau_2,f_2}(t_n')\right)^2.
\end{split}
\end{equation}
By replacing $\bm\zeta$ in $\LIFF$ by $\tilde{\bm{g}}^\text{w}_{\tau,f}$ and taking the limit as $\sigma^2 \rightarrow 0$, we see that this metric is equivalent to $\LIFF$ in the absence of noise,
$$
\lim_{\sigma^2\rightarrow 0} -2\sigma^2 \LIFF (\tilde{\bm{g}}^\text{w}_{\tau_1,f_1};\tau_2,f_2 ) = \mathcal{D}(\tau_1,f_1,\tau_2,f_2).
$$

\begin{figure}
\begin{subfigure}[b]{0.5\linewidth}
\centering
\includegraphics[width=\textwidth]{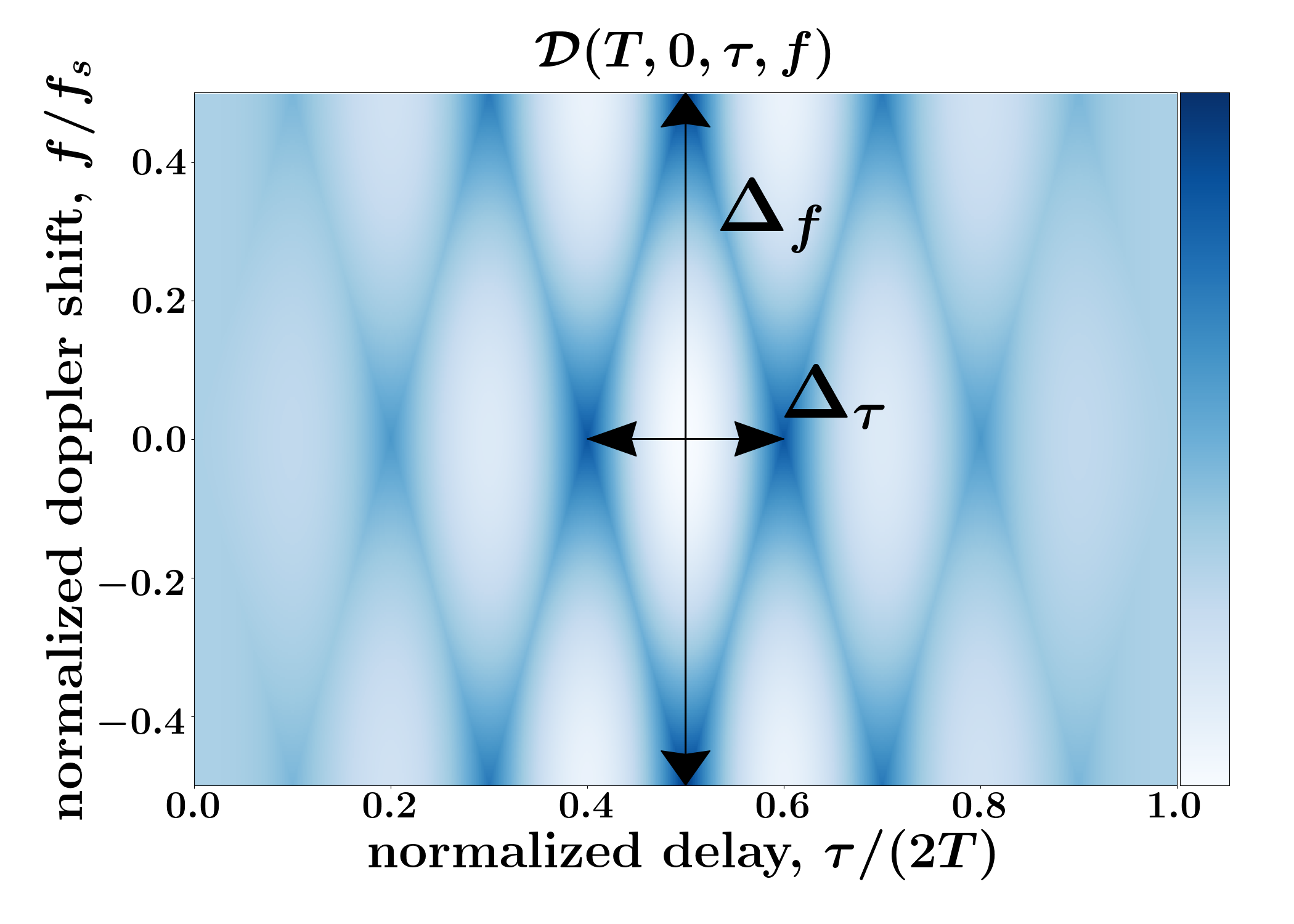}
            \vspace{-3mm}
        \caption{Triangular Modulation}\label{fig:trigeo}
    \end{subfigure}%
    \begin{subfigure}[b]{0.5\linewidth}
        \centering
        \includegraphics[width=\textwidth]{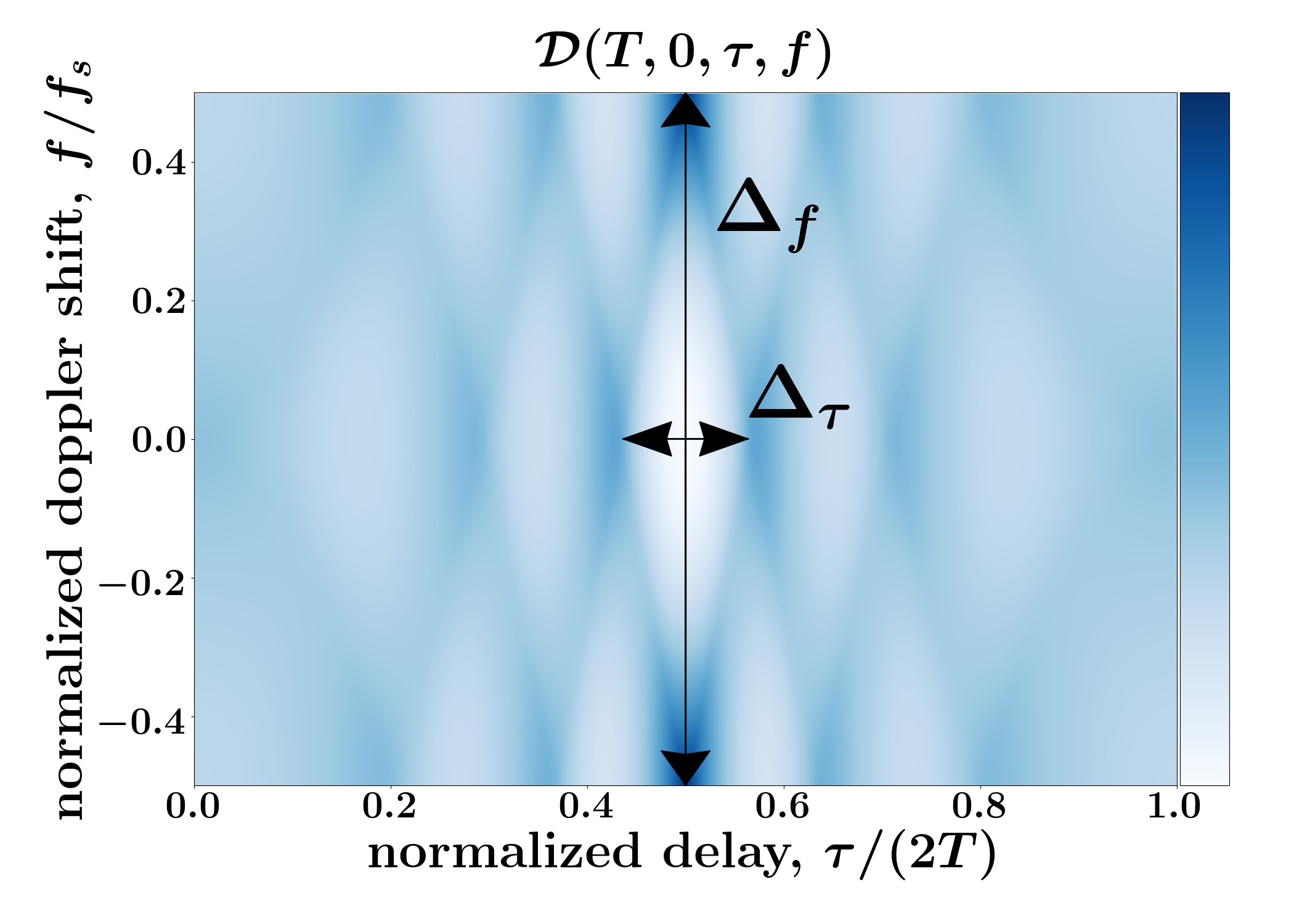}
        \vspace{-3mm}
        \caption{Sinusoidal Modulation}\label{fig:singeo}
        \end{subfigure}
\caption{Plot of $\mathcal{D}^2(T,0,\tau,f)$
as $(\tau,f)$ is varied
for (a) triangular modulation and (b) sinusoidal modulation. 
The whiter color indicates lower $\mathcal{D}^2$, and the bluer color indicates larger distance. 
The global optimum is located in the center. 
We can see the local optima are making the rhombus structures. 
For triangular modulation, the rhombuses have equal width, whereas for sinusoidal modulation the rhombus becomes wider as one moves away from the global optimum.}
\label{fig:geodesic_contour}
\end{figure}

\begin{figure*}
\centering
\includegraphics[width=0.85\linewidth]{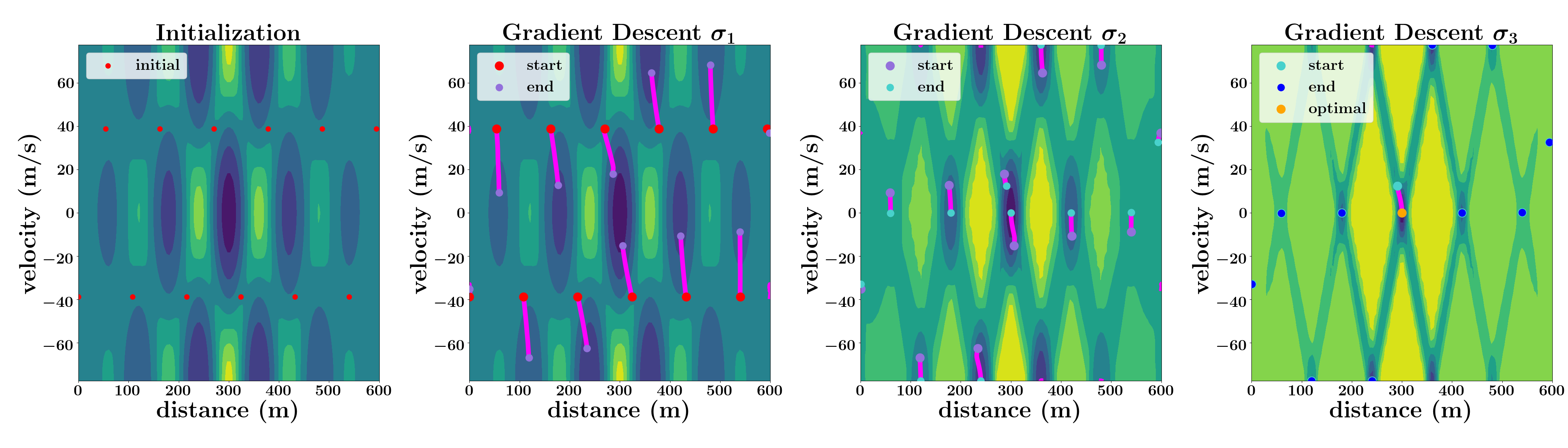}
\caption{The leftmost plot shows an example contour plot of relaxed $\LIFF$ and initial points for gradient ascent.
Note the separation of local maxima is similar to \cref{fig:geodesic_contour}.
The plots to the right show the trajectories of the gradient descents while gradually unrelaxing
$\LIFF ~(\sigma_1 > \sigma_2 > \sigma_3 = \hat{\sigma})$.
One or more points converge to the global maximum, which becomes our final estimate.}
\label{fig:multiple_graident_descent}
\end{figure*}

\Cref{fig:geodesic_contour} shows
$\mathcal{D}^2(T,0, \tau_2,f_2)$
as a function of $(\tau_2,f_2)$
for both triangular and sinusoidal modulation.
We see
that even in the noiseless case, the objective is heavily non-convex.
Empirically, local maxima occur when the differences of some elements before the modulo operation are near the modulo boundaries.
For a given $(\tau_1, f_1)$,
local maxima occur near $(\tau_2, f_2)$ that satisfy 
\begin{equation}
\max_n \left| \tilde{{g}}^\text{w}_{\tau_1, f_1}(t_n') - \tilde{{g}}^\text{w}_{\tau_2,f_2}(t_n') \right| = \frac{1}{2},
\label{eq:localmax_condition}
\end{equation}
which implies that the non-convexity of the objective is related to aliasing. 
Smaller $\fs/B$ may cause aliasing to occur more times and increases the number of modes in the objective. 
Our proposed solution to this non-convex optimization is to run a gradient-based optimization algorithm from multiple initial points.
To reduce the complexity of the algorithm, we aim to minimize the number of initial points. 
Now that we know the landscape of the objective and approximately know where the local optima occur, we can construct a grid of initial points to efficiently find the global optimum.

\subsection{Number and Placement of Initial Points}

To see how many initial points we need,  consider \cref{fig:geodesic_contour} again.
The contour map creates a series of rhombus-shaped basins of attraction. 
To achieve convergence to the global optimum from
at least one of the initial points,
one initial point must lie inside the rhombus that contains the global optimum. 
Thus, the required resolution of the grid of initial points is influenced by the size of
this rhombus.

Based on \eqref{eq:localmax_condition}, we can find the width $\Delta_\tau$ and height $\Delta_f$ of the rhombus that contains the global optimum. For triangular modulation,
\begin{align}
\Delta_\tau = \frac{\fs T}{B},
\qquad
\Delta_f = \fs. \label{eqn:trimod_rhombus}
\end{align}
For sinusoidal modulation, 
\begin{align}
\Delta_\tau \approx \arcsin\!\left( \frac{\fs}{2B}\right) \frac{4T}{\pi},
\quad
\Delta_f = \fs. \label{eqn:sinmod_rhombus}
\end{align}
For arbitrary modulation, we can construct $\mathcal{D}^2(T,0, \tau_2,f_2)$ offline to approximate  $\Delta_\tau$ and $\Delta_f$.

Our actual objective function $\LIFF$ is smoother compared to $\mathcal{D}^2$ because of $\sigma^2 > 0$, and the stochasticity of $\bm\zeta$ could perturb the relative placements of local optima.
When optimizing $\LIFF$, we generate initial points in a lattice structure where the separation in frequency is $\Delta_f$ and the separation in delay is $\gamma \Delta_\tau$, where $\gamma < 1$ is a parameter that prevents the failure of convergence to the global optimum due to stochastic perturbations.
In our experiments, the stochasticity of $\bm\zeta$ has little effect on the structure, and $\gamma$ could be quite high; we used $\gamma=0.9$.
An example of the placement of initial points is shown in the leftmost plot of \cref{fig:multiple_graident_descent}.
Algorithm~\ref{alg:ifreg} summarizes the procedure of our proposed estimator.

\begin{algorithm}
    \caption{IF Regression Estimator}
    \label{alg:ifreg}
    \begin{algorithmic}[1]
        \renewcommand{\algorithmicrequire}{\textbf{Input:}}
        \renewcommand{\algorithmicensure}{\textbf{Output:}}
        \Require $u(t_n)$, $v(t_n)$, $L$, $K$, $\gamma$
        \Ensure  $\hat{d}, \hat{v}$
        \State compute $\zeta_n$ from $u(t_n)$ (\cref{eqn:zeta})
        \State compute $\SNRetaHat$ from $u(t_n)$ and $v(t_n)$ (\cref{eqn:snrhat})
        \State compute $\hat{\sigma}^2$ (\cref{eqn:sigmahat})
        \State  generate $\{{\tau}_i, {f}_i\}_i$ in rhombus structure (\cref{eqn:trimod_rhombus,eqn:sinmod_rhombus})
        \State optimize $\LIFF(\bm \zeta; {\tau}_i, {f}_i)$ over ${\tau}_i, {f}_i$ for each $i$
        \State $(\hat{\tau}, \hat{f}) \leftarrow \argmax_{({\tau}_i, {f}_i)} \LIFF( \bm\zeta; \tau_i,f_i )$
        \State $\hat{\tau} \leftarrow \hat{\tau}\text{ mod }2T$
        \State $\hat{f} \leftarrow\Omega_{\fs}(f)$ \\
        \Return $\hat{\tau}, \hat{f}$
    \end{algorithmic}
\end{algorithm}

Since $\sigma$ can be quite small ($\sigma \ll 1$), numerical underflow may prevent convergence.
In \cref{fig:multiple_graident_descent}, we show an example strategy for a gradient descent algorithm, where we relax $\LIFF$ by setting $\sigma^2$ to a value larger than the estimated value as described in \cref{sec:variance_approx}.
We gradually decrease $\sigma^2$
as we iterate,
ultimately using the estimated variance $\hat{\sigma}^2$. 
Our implementation of the IFF method can be found in 
\href{https://github.com/Goyal-STIR-Group/FMCW-Above-Nyquist.git}{https://github.com/Goyal-STIR-Group/FMCW-Above-Nyquist.git}.

\section{Numerical Results}

\label{sec:num_results}

We test our proposed processing approaches through a number of simulations.
Assuming
the surface area of the reflected object is larger than the beamwidth, 
we follow the two-way lidar equation~\cite{kim_thesis_2020} to determine the appropriate received power given a distance $d$: 
\begin{equation}
    \PRX = \PTX \frac{\RPD A \rho \cos (\phi)}{\pi d^2} \exp\!\left(-{\textstyle\int \alpha(d) \diff d}\right).
\end{equation}
In all our simulations, we neglect atmospheric absorption
(i.e., $\alpha(d) = 0$)
and assume the target is fronto-parallel ($\phi = 0$).
We set the target reflectivity $\rho = 0.01$,
photodetector responsivity $\RPD=1$,
transmit power $\PTX= \SI{1}{\milli\watt}$, and
receiver aperture area $A = \SI{1}{\milli\meter^2}$. 
The laser linewidth is set to \SI{100}{\kilo\hertz}.

For the IFF method, we used $K=2$ for evaluating $\LIFF$ and $\gamma=0.9$ for controlling the spacing of the initial points. 
We use the Broyden--Fletcher--Goldfarb--Shanno (BFGS) algorithm~\cite{broyden_1970_bfgs,fletcher_1970_bfgs,goldfarb_1970_bfgs,shanno_1970_bfgs} for optimization from each initial point.

\begin{figure*}
\centering
\begin{subfigure}[b]{\linewidth}
    \centering
    \includegraphics[width=0.9\textwidth]{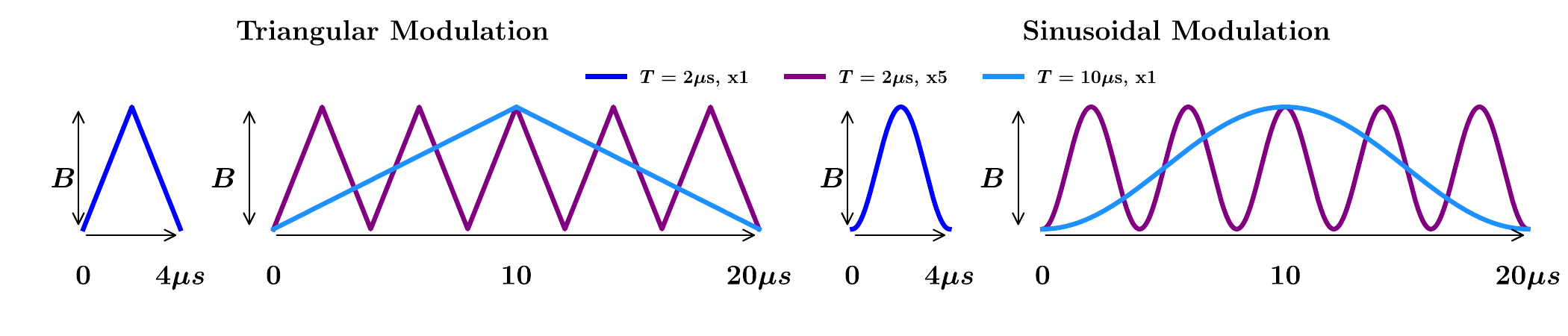}
    \vspace{-5mm}\caption{  Acquisition Settings}\label{fig:desta}
\end{subfigure}\\
\vspace{5mm}
\centering
\begin{subfigure}[b]{0.24\linewidth}
    \centering
    \includegraphics[width=0.99\textwidth]{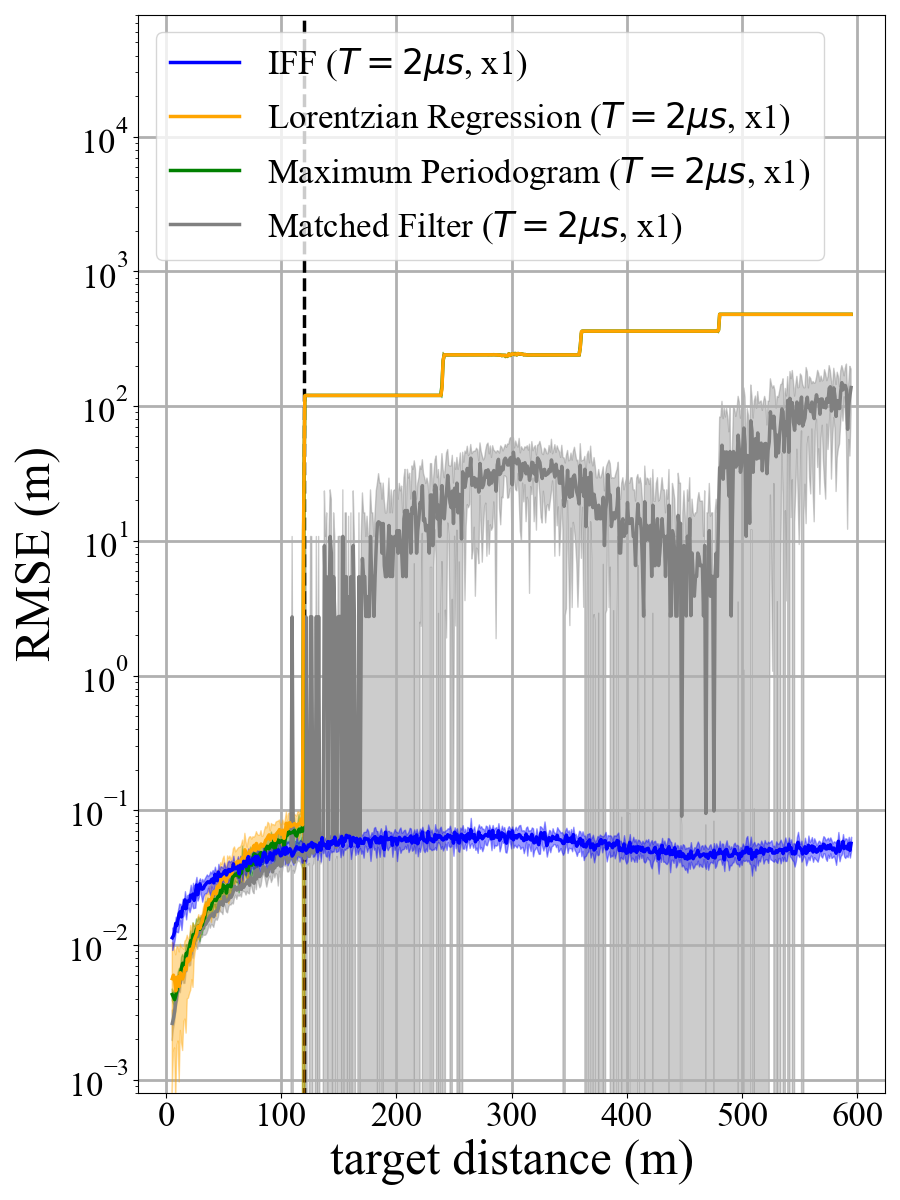}
    \caption{  Triangular (\SI{4}{\micro\second} observation)}\label{fig:destb}
\end{subfigure}%
\centering
\begin{subfigure}[b]{0.24\linewidth}
    \centering
    \includegraphics[width=0.99\textwidth]{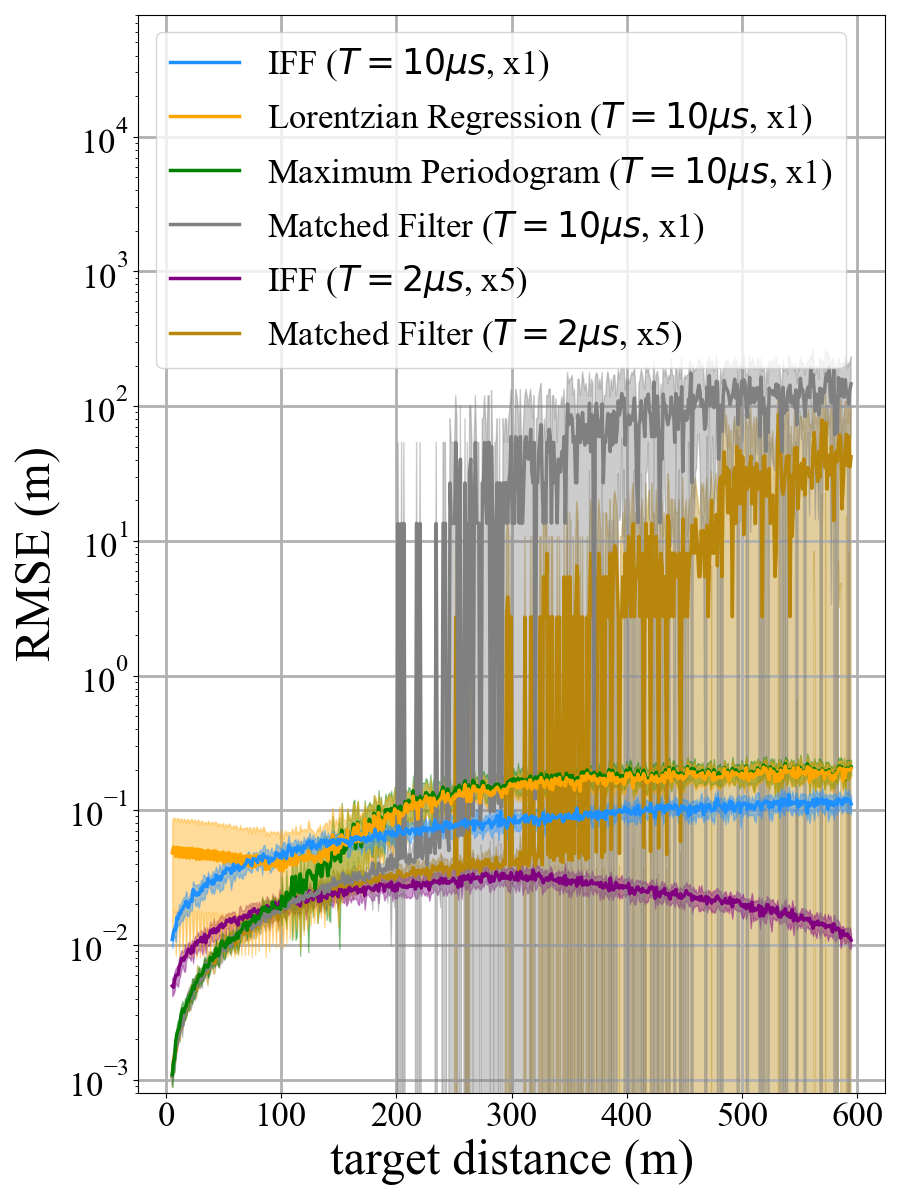}
    \caption{  Triangular (\SI{20}{\micro\second} observation)}\label{fig:destc}
\end{subfigure}%
\centering
\begin{subfigure}[b]{0.24\linewidth}
    \centering
    \includegraphics[width=0.99\textwidth]{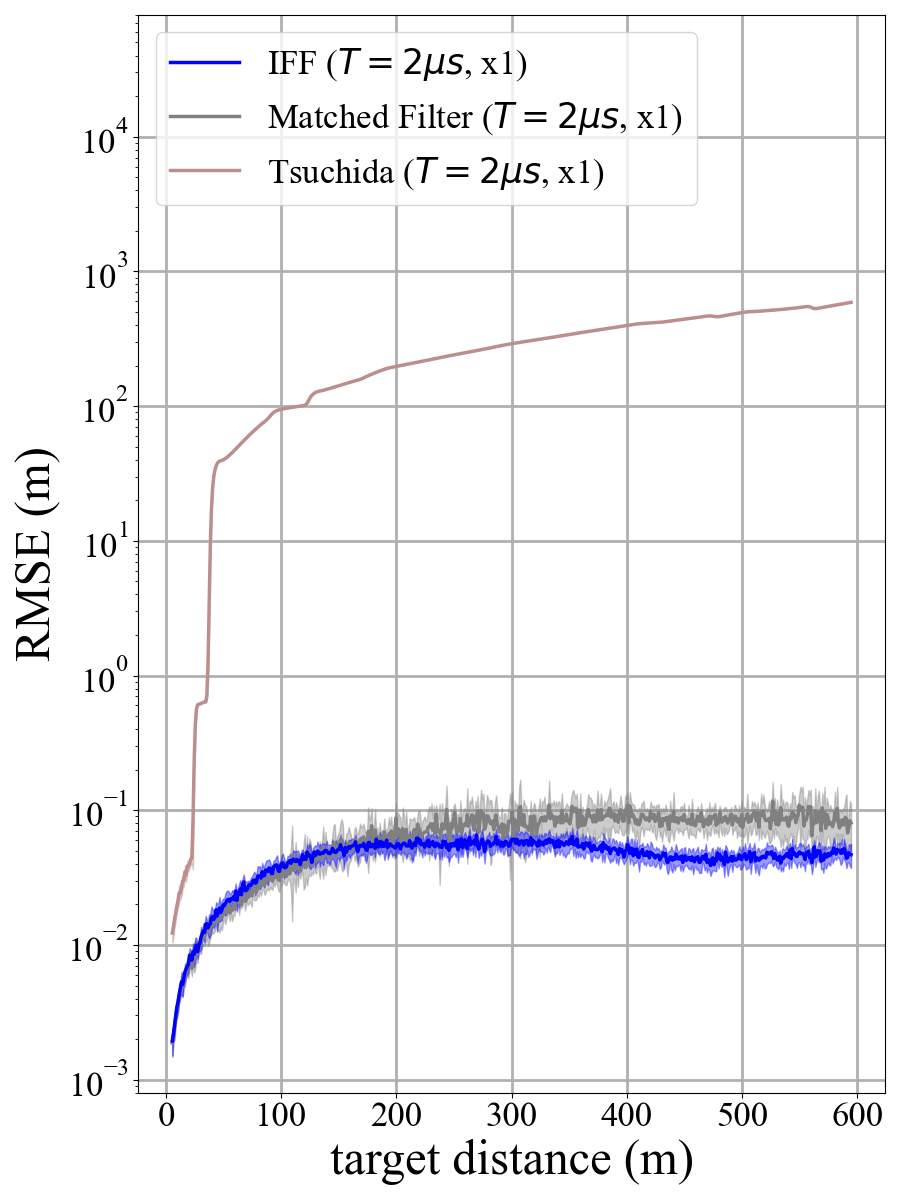}
    \caption{  Sinusoidal (\SI{4}{\micro\second} observation)}\label{fig:destd}
\end{subfigure}
\centering
\begin{subfigure}[b]{0.24\linewidth}
    \centering
    \includegraphics[width=0.99\textwidth]{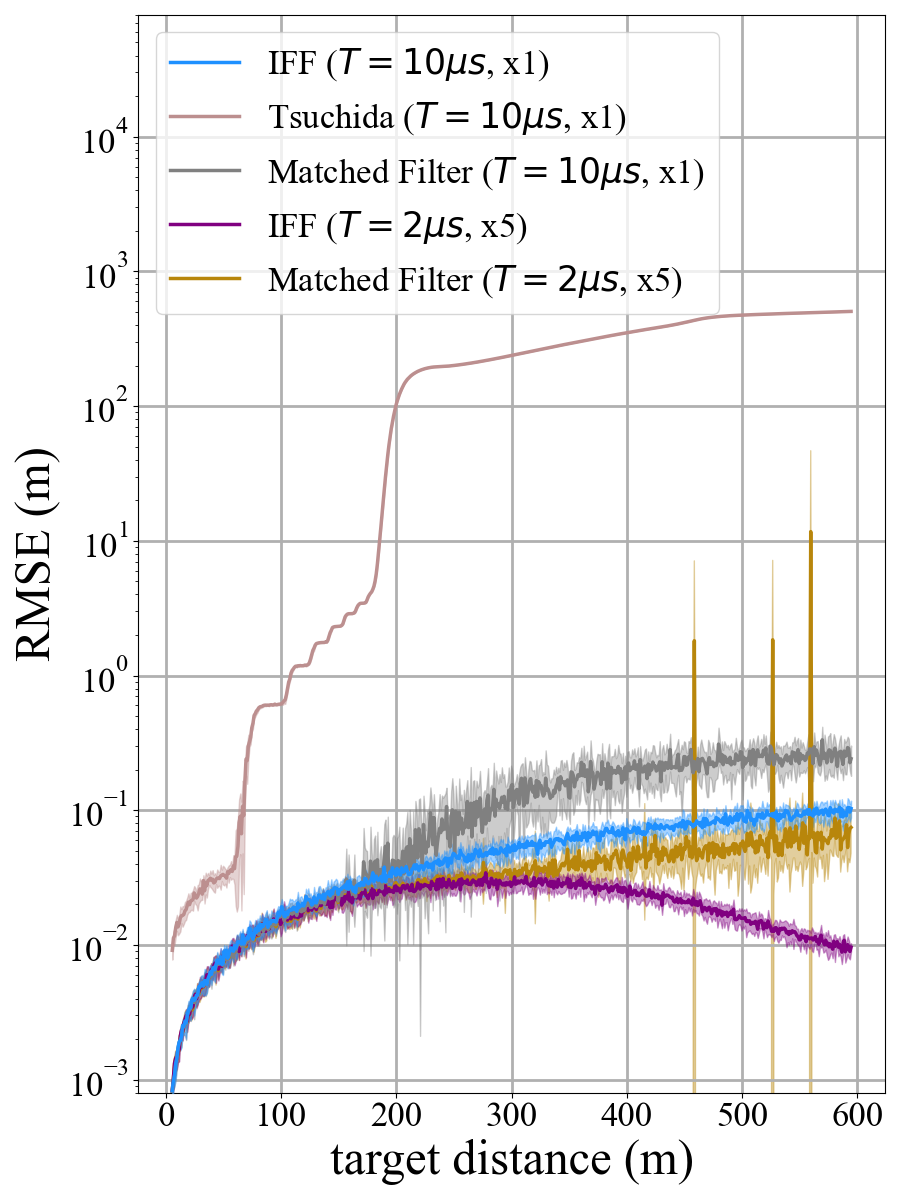}
    \caption{  Sinusoidal (\SI{20}{\micro\second} observation)}\label{fig:deste}
\end{subfigure}
\caption{
Simulation results of distance-only estimation (zero velocity), with $\fs = \SI{200}{\mega\hertz}$ and $B = \SI{500}{\mega\hertz}$. 
We illustrate in~\subref{fig:desta} the different modulation functions and number of repetitions used for the experiments.
The results with triangular modulation with $T = \SI{2}{\micro\second}$ in~\subref{fig:destb} show conventional methods are clearly limited by the unambiguous range.
In \subref{fig:destc}, $T$ is adjusted to $\SI{10}{\micro\second}$, and the IFF method still outperforms the conventional methods.
The RMSE of the IFF method can be improved further without adjusting $T$ by observing 5 consecutive periods that equate to the same number of samples. Similar results are shown for sinusoidal modulation with $T = \SI{2}{\micro\second}$ in~\subref{fig:destd} and with $T = \SI{10}{\micro\second}$ in~\subref{fig:deste}, although the matched filter performance is much better for sinusoidal modulation than triangular modulation.
Shaded areas indicate the standard deviation of the RMSE curves.}

\label{fig:dest}
\end{figure*}

\subsection{Comparison Methods}

We briefly review the implementation of the methods used for comparison.

\subsubsection{Maximum Periodogram}
We implement the maximum periodogram estimator~\eqref{eq:max_periodogram} via a two-step process.
We first initialize our estimates based on the maximum peak observed in the modulus of DFT\@. 
We then refine the result to get continuous-valued beat frequency estimates using Brent's method~\cite{brent_algorithms_1973}.

\subsubsection{Lorentzian Fitting}
Kim et al.~\cite{kim_optimal_2018} account for spectral broadening due to phase noise by fitting a Lorentzian function to the magnitude of the DFT of the signal.
We compute an initial frequency estimate using the maximum periodogram method as described above.
We then refine that estimate by nonlinear least-squares fitting of a Lorentzian function to the power spectral density using the BFGS algorithm.

\subsubsection{Tsuchida's Method}
Tsuchida~\cite{tsuchida_freqavg_2019} proposes a method to sequentially estimate the Doppler shift and time delay from sinusoidal modulation using the instantaneous frequency and phase.
We implement Tsuchida's method as
\begin{align*}
    \hat{f} &= \frac{f_s}{N-1}\sum_{n=0}^{N-2}\xi_i, \\
    \hat{\tau} &= \frac{\frac{1}{N}\sum_n\left|(\angle u(t_n))_{\text{unwrap}}-2\pi\hat{f}t_n\right|}{\frac{\pi}{T}\int_0^{2T} |a(t)|dt},
\end{align*}
where
$\angle u(t))_{\text{unwrap}}$ is the unwrapped phase of the received signal $u(t)$. 

\subsection{Distance Estimation}

We first discuss our simulation results for distance estimation with zero velocity ($f=0$), shown in \cref{fig:dest}. 
We consider both triangular and sinusoidal modulations over either $1$ or $5$ chirp periods, as illustrated in \cref{fig:desta}. 
For the results shown in \cref{fig:destb,fig:destc,fig:destd,fig:deste}, we generate complex-valued measurements at distances ranging from \SI{5}{\meter} to \SI{599}{\meter} with \SI{0.1}{\meter} increments. 
For each distance, we generate $20$ measurements and plot the root mean-squared error (RMSE) and the sample standard deviation of the squared error.

In \cref{fig:destb}, we use triangular modulation with  
$T=\SI{2}{\micro\second}$, $B=\SI{500}{\mega\hertz}$, $\fs = \SI{200}{\mega\hertz}$, and the observation window is set to 
$\SI{4}{\micro\second}$ (one period of FMCW signal that includes an up-chirp and a down-chirp). This corresponds to $N=800$ samples in one measurement.
With complex measurements, both maximum periodogram and Lorentzian fitting can only resolve distances up to $\Frac{(c T \fs)}{(2B)}$, which in this case is \SI{120}{\meter}.
Instantaneous frequency fitting
can resolve up to
an unambiguous range of $(c/2)(2T)=cT$, which is \SI{600}{\meter}.
These limits are confirmed in \cref{fig:destb}, as the RMSE for the conventional algorithms significantly increases beyond \SI{120}{\meter}, whereas the IFF algorithm achieves $\text{RMSE}<\SI{10}{\centi\meter}$ up to \SI{600}{\meter}.
This demonstrates the first strength of our proposed approach, which is much \textbf{\textit{larger unambiguous range}}. 
While matched filtering can handle arbitrary waveforms as well as aliasing, the undersampled measurement ($f_s < B$) causes spurious sidelobe peaks in the cross-correlation.
Large errors occur when phase noise causes higher sidelobes than the mainlobe peak.
Further discussion of this phenomenon follows in \cref{subsec:mf_peaks}.

In \cref{fig:destc}, 
we evaluate the benefit of our methods even when aliasing does not occur.
We modify the triangular modulation parameters by setting $T=\SI{10}{\micro\second}$ with unchanged bandwidth,
and the observation window is set to $\SI{20}{\micro\second}$ (one period of FMCW signal).
We now use $4000$ samples, 
and the maximum unambiguous distance for CBF-based methods increases to $\SI{600}{\meter}$. 
The maximum periodogram and matched filtering methods have the best performance at short distances, for which phase noise is negligible, 
whereas the IFF method has relatively weaker performance
due to inaccurate modeling of the noise correlation from phase differentiation. 
We discuss in \cref{sec:analysis} how this inaccurate modeling affects the theoretical performance of the IFF method. 
As the distance increases, however, the CBF segments become shorter and the accuracy of maximum periodogram drops, while matched filtering still suffers from large sidelobes. 
The IFF method achieves \textbf{\textit{better sample efficiency}} over CBF-based methods through more accurate modeling of non-CBF segments and is more robust against the phase noise. 

A benefit of tolerating aliasing in the measurement is that there was no need to adjust $T$ to resolve distances up to $\SI{600}{\meter}$.
In \cref{fig:destc}, we also show the result of the IFF method for $T = \SI{2}{\micro\second}$ but with 5 consecutive periods over the same $\SI{20}{\micro\second}$ observation window. Because of the higher chirp rate, the observed signal is more sensitive to the distance. 
Using the IFF method for FMCW signal with this set of parameters ($T=\SI{2}{\micro\second}$, $5$ chirps) decreases the RMSE even further. 
This result demonstrates that the IFF method does not have the same range/resolution tradeoffs as CBF-based methods.

\begin{figure*}[th!]
\begin{subfigure}[b]{\linewidth}
    \centering
    \includegraphics[trim={0 30mm 0 32mm}, clip, width=0.95\linewidth]{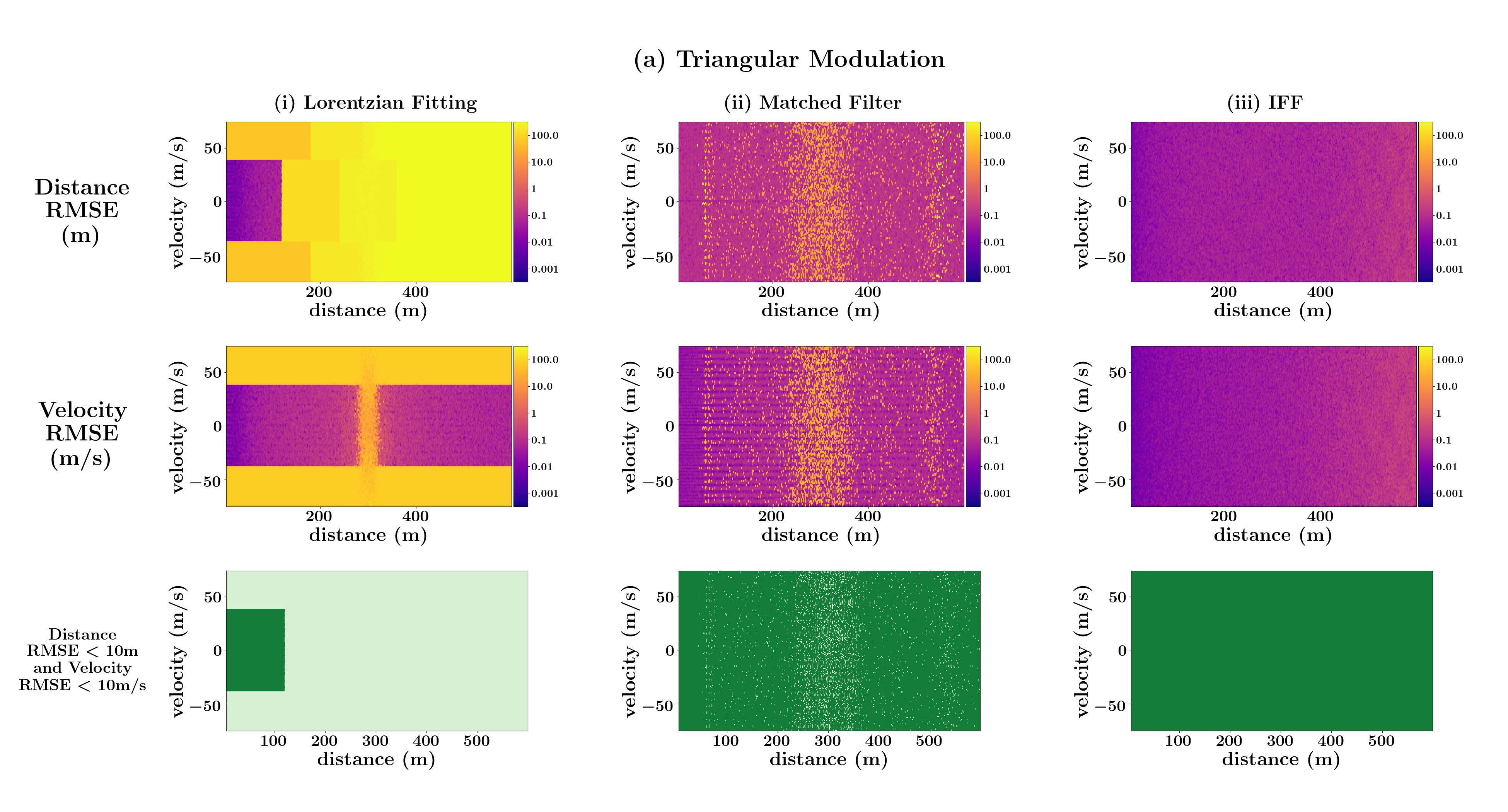}
    {\phantomsubcaption\label{fig:dvest_a}}
\end{subfigure}\\
\begin{subfigure}[b]{\linewidth}
    \centering
    \includegraphics[trim={0 30mm 0 10mm}, clip, width=0.95\linewidth]{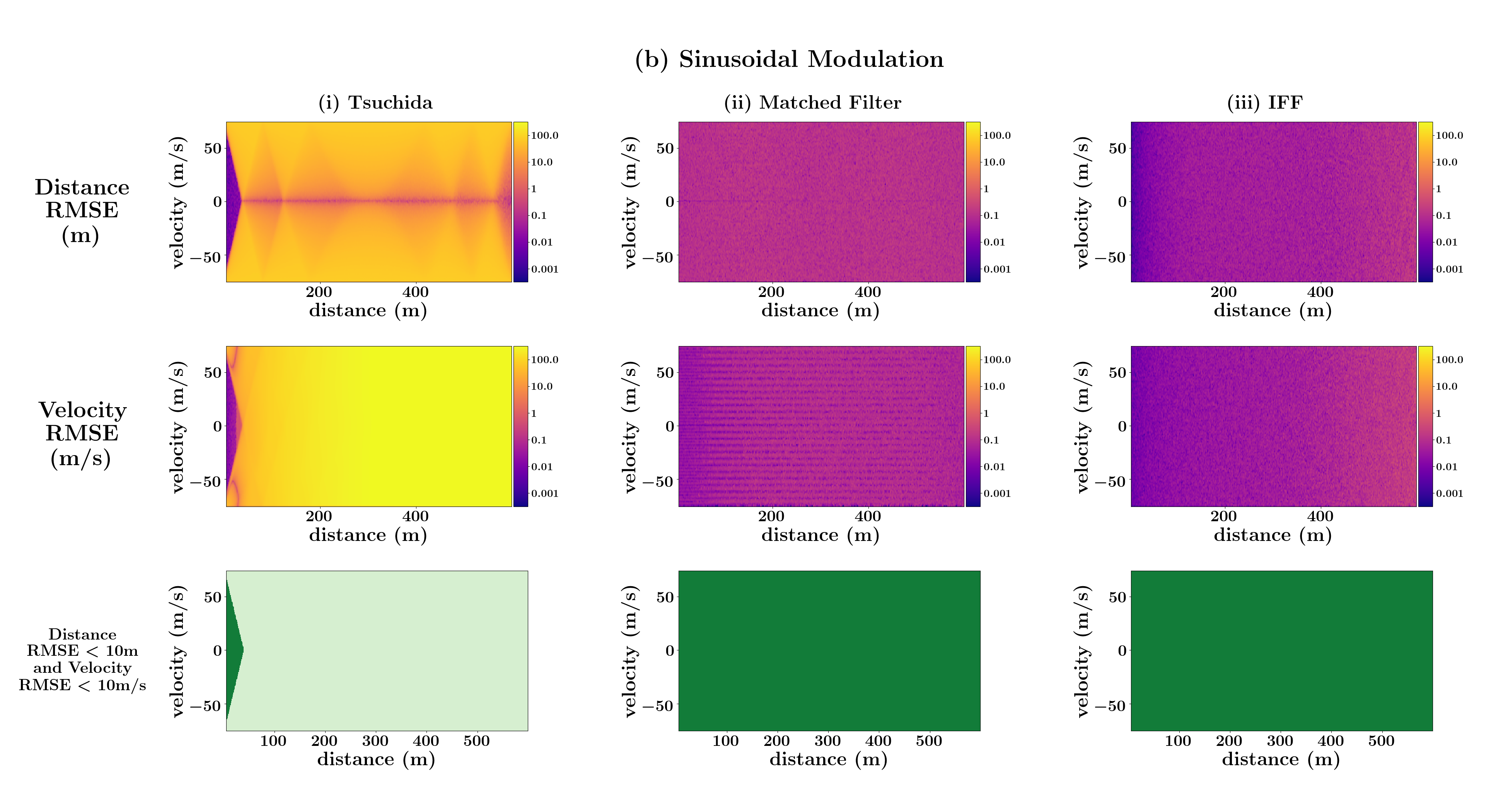}
    {\phantomsubcaption\label{fig:dvest_b}}
\end{subfigure}
    \caption{Simulation results for joint distance--velocity estimation with triangular modulation (a) and sinusoidal modulation (b).
    In (a) we compare Lorentzian fitting (i) and matched filtering (ii) to our IFF method (iii).
    In (b) we compare Tsuchida's method (i) and matched filtering (ii) to our IFF method (iii).
    For each modulation, we use $T = \SI{2}{\micro\second}$, $B = \SI{5}{\giga\hertz}$, and $\fs = \SI{2}{\giga\hertz}$.
    We show the distance RMSE of each method in the top row and the velocity RMSE in the middle row.
    In the bottom row, we highlight in dark green areas that have both distance RMSE less than  \SI{10}{\meter} and velocity RMSE below \SI{10}{\meter/\second}; otherwise, light green is shown.
    }
    \label{fig:dvest}
\end{figure*}

To demonstrate \textbf{\textit{applicability to nonlinear frequency modulation}} with our IFF method, we show distance estimation with sinusoidal modulation in \cref{fig:destd,fig:deste}.
For sinusoidal modulation, we compare our results against two methods:
matched filtering and Tsuchida's method~\cite{tsuchida_freqavg_2019}. 
Similar to triangular modulation, we show the results over three sets of parameters: 
($T= \SI{2}{\micro\second}$, $B=\SI{500}{\mega\hertz}$, $\SI{4}{\micro\second}$ observation window, 800 samples),
($T=\SI{10}{\micro\second}$, $B=\SI{500}{\mega\hertz}$, $\SI{20}{\micro\second}$ observation window, 4000 samples), and
($T= \SI{2}{\micro\second}$, $B=\SI{500}{\mega\hertz}$, $\SI{20}{\micro\second}$ observation window, 4000 samples).
The sampling rate is fixed to $\SI{200}{\mega\hertz}$ as before.
Tsuchida's method is a simple and fast method of computing delay and Doppler shift from a sinusoidal modulation. 
However, it requires phase unwrapping, which fails when the IF is beyond the Nyquist frequency at any given point. 
Unlike in the results for triangular modulation, matched filtering performance is significantly better due to much lower sidelobes, making sinusoidal modulation a more suitable waveform for matched filtering when the measurement signal is undersampled.
However, matched filtering occasionally still fails while our IFF method is stable. We discuss the reason for
this stability in \cref{subsec:mf_peaks}.

As also observed for the triangular modulation,
the best overall performance is achieved by the IFF method using multiple short chirps over a longer acquisition window.
Both the matched filtering and IFF method achieve similar improved performance at short distances for the sinusoidal modulation.

\subsection{Joint Distance--Velocity Estimation}

We next evaluate simulation results for the joint estimation of distance and velocity, which are shown in \cref{fig:dvest}. 
We compare the IFF method against Lorentzian fitting~\cite{kim_optimal_2018} and matched filtering for triangular modulation in \cref{fig:dvest_a},
and against Tsuchida's method~\cite{tsuchida_freqavg_2019} and matched filtering for sinusoidal modulation in \cref{fig:dvest_b}.
In addition to showing RMSE for distance and velocity estimation in the first two rows of each subfigure,
the last row shows the region of distance--velocity pairs that result in RMSE less than \SI{10}{\meter} in distance and \SI{10}{\meter/\second} in velocity, which are marked in dark green for each method.
This region with acceptable performance for Lorentzian fitting is significantly smaller than for the IFF method, and its size is consistent with the unambiguous space we stated earlier in \cref{fig:first_d}. 
Tsuchida's method requires unwrapping the instantaneous phase, but
unwrapping fails when the absolute value of IF exceeds $\Frac{\fs}{2}$ at any point.
For this reason, Tsuchida's method has a much smaller operating region with acceptable RMSE than the IFF method.
For the matched filter, we see results similar to the distance-only estimation (\cref{fig:dest}). 
Sinusoidal modulation is much more stable for matched filtering, but the IFF method has better accuracy overall.

\subsection{Computational Complexity}
We provide in \cref{tab:avg_comp_time_tri} and \cref{tab:avg_comp_time_sin} the average runtime of the methods for both distance-only and joint distance--velocity estimation. 
Measurements are generated using the same parameters as in \cref{fig:desta,fig:destc}, and the runtime is averaged over various distance and velocity combinations. 
Computations were performed using an AMD Ryzen 5 5600X 6-core CPU. 
The CBF-based methods (maximum periodogram and Lorentzian fitting) and Tsuchida's method have no additional complexity for joint distance--velocity estimation. 
While the RMSE obtained with the matched filter is occasionally competitive with the IFF method, the matched filter computation does not scale well with the added velocity dimension ($O(N) =N^2\log(N)$). 
Meanwhile, the IFF method's complexity increases by only a factor of 4 for joint estimation: a $2\times$ increase in the number of parameters and a $2\times$ increase in the number of initial points.
The IFF method also has similar distance-only computational complexity to Lorentzian fitting but can handle arbitrary modulations and achieve a larger unambiguous parameter space.

\begin{table}
\centering
\caption{Average computational time for triangular modulation (ms)}
\label{tab:avg_comp_time_tri}
\begin{tabular}{@{}lr@{\,}c@{\,}lr@{\,}c@{\,}l@{}}
\hline
\textbf{Method}
& \multicolumn{3}{@{}c}{\textbf{Distance only}}
& \multicolumn{3}{@{}c@{}}{\textbf{Distance and velocity}}\\
\hline
Maximum periodogram &  0.64 & $\pm$ & 0.06 & 0.66 & $\pm$ & 0.08 \\
Lorentzian fitting & 27.49 & $\pm$ & 6.47 & 28.11 & $\pm$ & 6.44 \\
Matched filtering & 1.49 & $\pm$ & 0.11 & 220.21 & $\pm$ & 23.26 \\
IFF & 25.17 & $\pm$ & 7.22 &  90.42 & $\pm$ & 24.97\\
\hline
\end{tabular}
\end{table}

\begin{table}
\centering
\caption{Average computational time for sinusoidal modulation (ms)}
\label{tab:avg_comp_time_sin}
\begin{tabular}{@{}lr@{\,}c@{\,}lr@{\,}c@{\,}l@{}}
\hline
\textbf{Method}
& \multicolumn{3}{@{}c}{\textbf{Distance only}}
& \multicolumn{3}{@{}c@{}}{\textbf{Distance and velocity}}\\
\hline
Tsuchida          &  0.17 & $\pm$ & 0.03 &   0.18 & $\pm$ & 0.02 \\
Matched Filtering &  1.01 & $\pm$ & 0.10 & 207.75 & $\pm$ & 16.46 \\
IFF & 27.32 & $\pm$ & 1.88 &  84.78 & $\pm$ & 5.21 \\
\hline
\end{tabular}
\end{table}

\subsection{Spurious Sidelobe Peaks for the Matched Filter}
\label{subsec:mf_peaks}

\begin{figure}
\begin{subfigure}[b]{\linewidth}
    \centering
    \includegraphics[width=\textwidth]{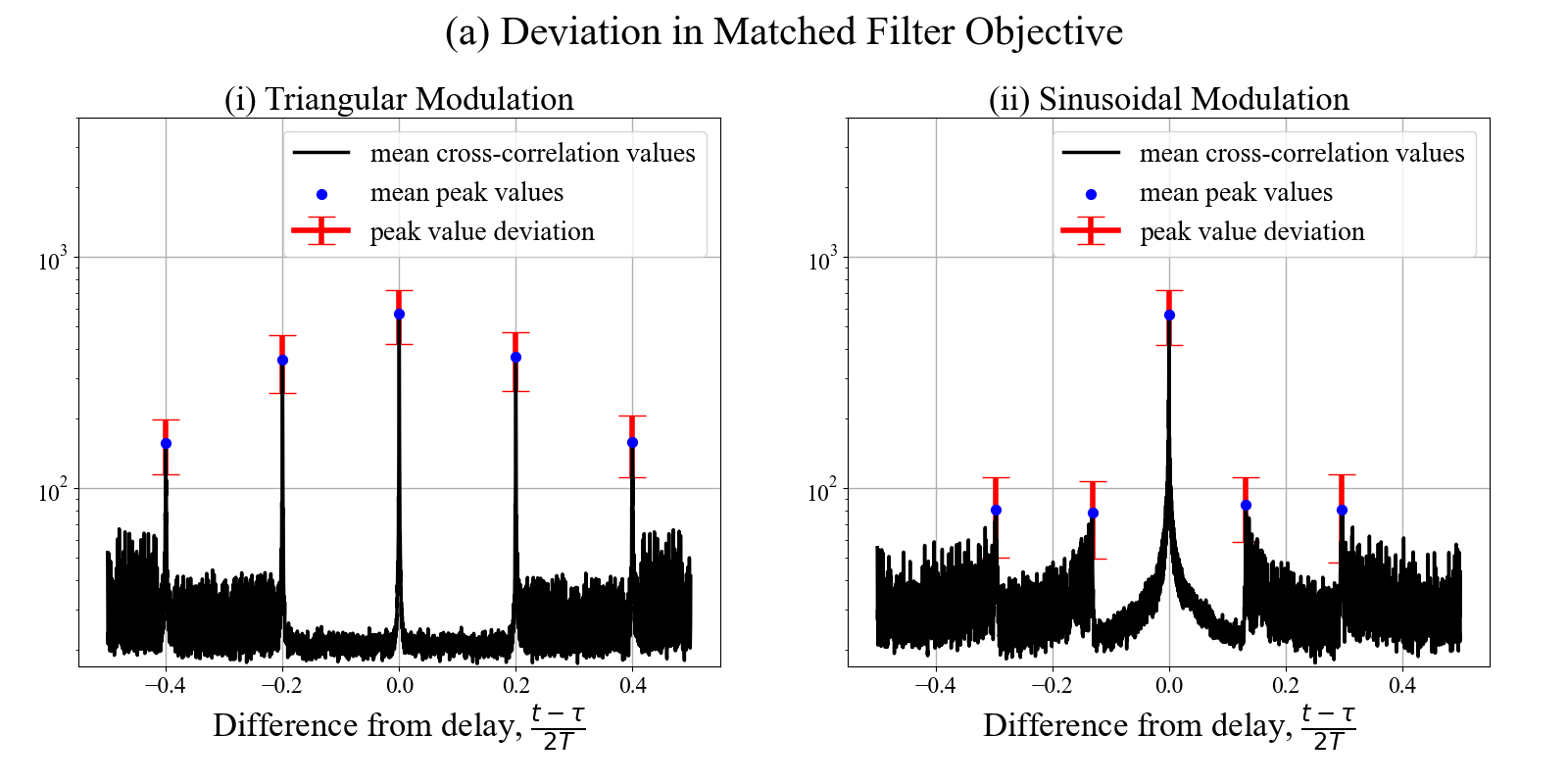}
    \vspace{-3mm}
    {\phantomsubcaption\label{fig:mf_dev}}
\end{subfigure}\\
\begin{subfigure}[b]{\linewidth}
    \centering\includegraphics[width=\textwidth]{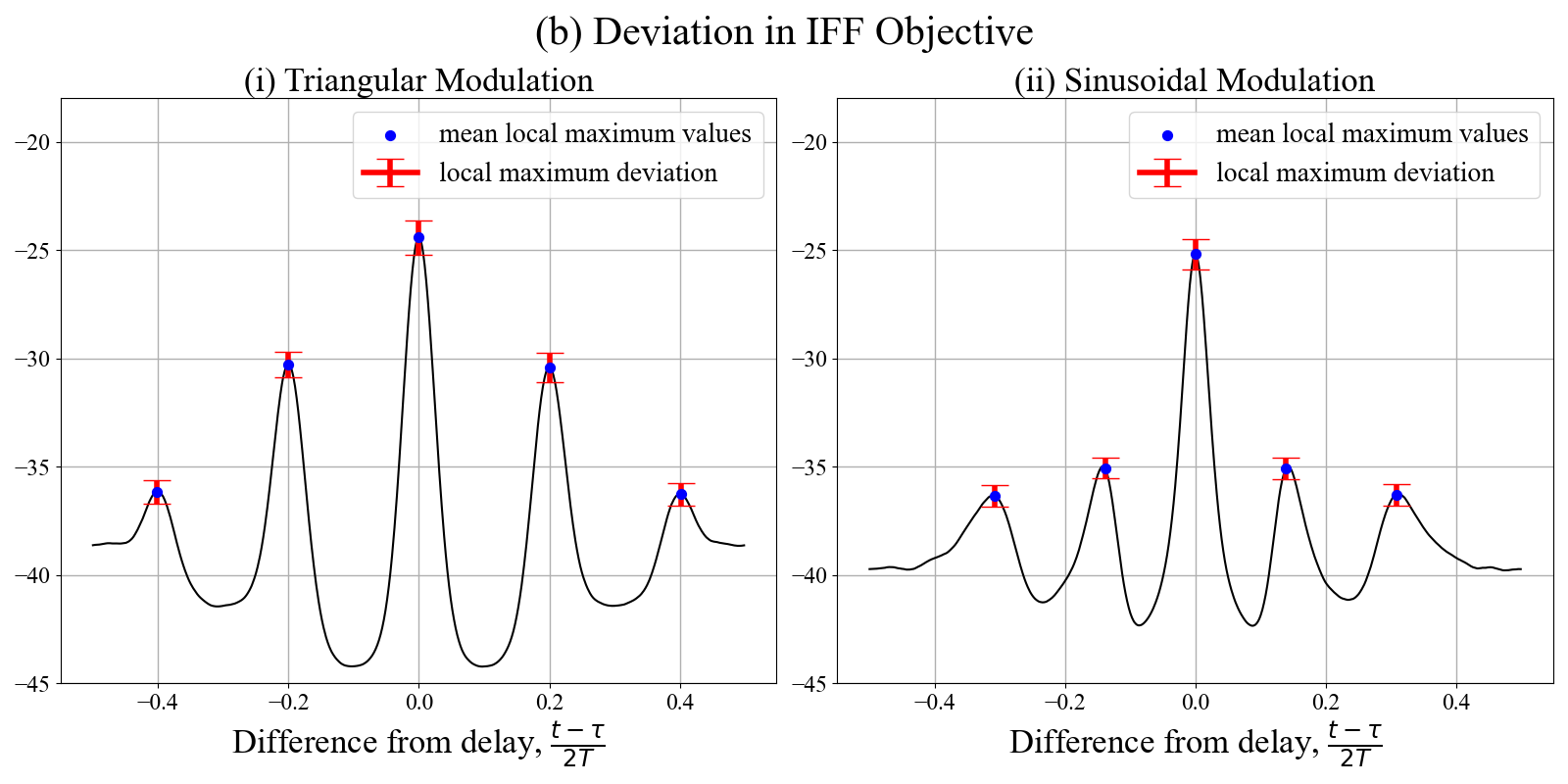}
    \vspace{-3mm}
    {\phantomsubcaption\label{fig:wn_dev}}
\end{subfigure}
\caption{(a) Mean of the modulus of matched filter response over 100 simulations of phase noise is shown. The center peak shows the response for the correct delay. For each peak, we illustrate the standard deviation with error bars. It is clear that triangular modulation is much more susceptible to phase noise than sinusoidal modulation.
(b) Mean of the wrapped normal likelihood function from \eqref{eq:LWN} over the same simulation is shown. Local maxima are located in similar locations as the peaks in the matched filter response. Similar to the matched filter response, the local maxima are relatively closer to the global maximum for triangular modulation than for sinusoidal modulation. Nonetheless, the standard deviation is not large enough to have large errors.
\label{fig:sidelobe_analyses}
}
\end{figure}

Revisiting \cref{fig:dest,fig:dvest},
it is interesting that the matched filter method has large RMSE for triangular modulation but not for sinusoidal modulation.
These trends can be understood by studying the matched filter response.
In \cref{fig:mf_dev}, we simulate phase noise and plot the mean values of the matched filter response.
For each modulation, we observe the peak in the center (main lobe) corresponding to the correct time delay as well as four other significant (sidelobe) peaks due to aliasing. 
The sidelobe peaks are much larger for triangular modulation than for sinusoidal modulation, and the standard deviations of the peaks further explain why a spurious peak is more likely to be confused for the correct peak in the case of triangular modulation.

In \cref{fig:wn_dev}, we show a similar analysis for the wrapped normal log-likelihood function~\eqref{eq:LWN}.
The locations of local maxima correspond to the local maxima of the matched filter response.
We again see that the global maximum is more separated from the other local maxima for the case of sinusoidal modulation as compared to triangular modulation.
The IFF method is more stable than the matched filter
because the deviations of the local maxima are small
relative to the gap between the global maximum and the other local maxima. 

\section{Theoretical Analysis of Estimation from Instantaneous Frequency Estimate $\zeta$}
\label{sec:analysis}

\subsection{Cram\'er--Rao Bound Analysis}
We aim to understand our numerical results further with the Cram\'er--Rao bound (CRB) and the misspecified CRB (MCRB). 
Here we focus on distance estimation of a stationary object.
As derived in \cref{apx:crb_delay}, the CRB for our problem is
\begin{equation}
\begin{split}
    &\text{CRB}_d =  \\
    & \left[ \left( \frac{\partial \tilde{\bm g}_d}{\partial d}\right)^T \bm \Sigma_d^{-1} \left( \frac{\partial \tilde{\bm g}_d}{\partial d}\right) + \frac{1}{2} \tr\!\left( \frac{\partial \bm\Sigma_d}{\partial d} \bm \Sigma_d^{-1}  \frac{\partial \bm\Sigma_d}{\partial d} \bm \Sigma_d^{-1} \right) \right]^{-1}.
\end{split}
\label{eq:CRB}
\end{equation}
where $\tilde{g}_d = (1/\fs) g_{\tau=2d/c,f=0}$, and $\Sigma_d$ is the covariance matrix of $\chi_n + \epsilon_n$, which is also a function of distance (equivalently time delay).
While the CRB is a more frequently used theoretical bound on the performance of an estimator, 
our estimator has a model misspecification, since it
assumes independence of the noise 
for tractable computation.
Under this scenario, the MCRB~\cite{vuong_mcrb_1986, fortunati_mcrb_2017} can be more informative for analyzing our estimator.
In \cref{apx:mcrb_delay}, we derive
the MCRB to be 
\begin{equation}
    \text{MCRB}_d = \frac{  \frac{\partial \tilde{\bm{g}}_d^T}{\partial d}  \bm\Sigma_d \frac{\partial \tilde{\bm{g}}_d}{\partial d} }{\left(\frac{\partial \tilde{\bm{g}}_d^T}{\partial d}  \frac{\partial \tilde{\bm{g}}_d}{\partial d}\right)^2}.
    \label{eqn:mcrb}
\end{equation}

\subsection{Covariance Matrix Approximation}
\label{sec:cov_mat_approx}

\begin{figure}
    \centering
    \includegraphics[width=\linewidth]{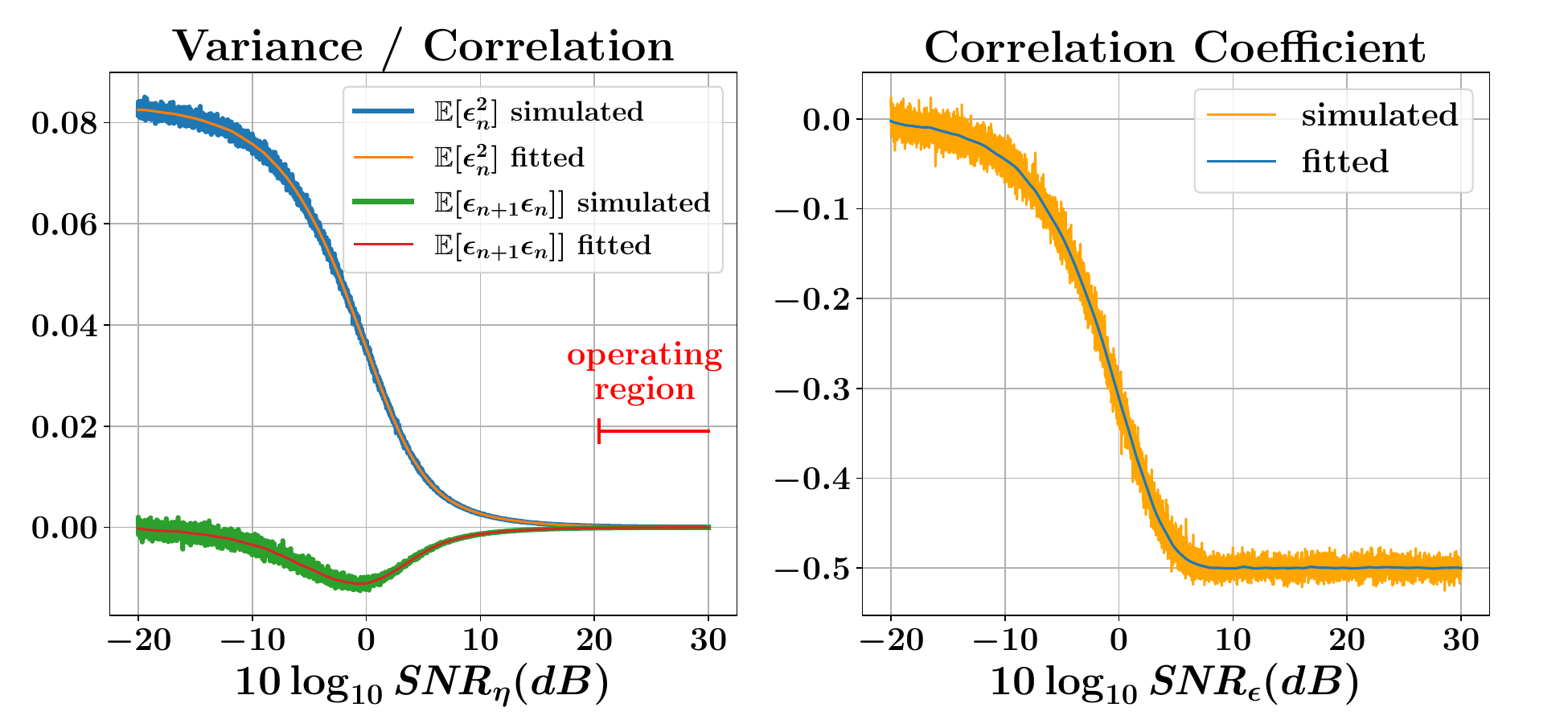}
    \caption{Noise analysis. The left plot shows the variance $\Var(\epsilon_{n})$ and correlation $\mathbb{E}[\epsilon_{n+1}\epsilon_{n}]$ as a function of $\SNReta$.
    The right plot shows the correlation coefficient $\mathbb{E}[\epsilon_{n+1}\epsilon_n]/\Var(\epsilon_{n})$.
    Note the correlation coefficient is $-1/2$ when $\SNReta$ is above \SI{10}{\dB} or so, but as $\SNReta$ decreases the correlation diminishes.
    The ``operating region'' denotes the range of $\SNReta$ values used in the numerical results of \cref{fig:dest,fig:dvest} for measurements up to \SI{600}{\meter}.}
    \label{fig:snrvar}
\end{figure}

Here we determine the correlation of noise in samples $\bm\Sigma_d$, which is used in CRB and MCRB expressions.
Previously in \eqref{eq:zeta_n_decomposition} we introduced a decomposition of $\bm\zeta$. 
Given that the autocorrelation for the phase noise $\xi(t)$ follows~\eqref{eq:phasenoiseautocorr} and the shot noise is uncorrelated,
i.e.,  $R_\eta(u) = \delta(u)$, then
the covariance matrix of the sum of the two noise sources is 
\begin{equation}  
    \left[\bm\Sigma_d \right]_{ij} = \begin{cases}
        \frac{1}{\pi\fs} L + p, & i=j; \\
        q,                           & |i-j| = 1; \\
        \frac{1}{2\pi} L (\tau -  \frac{\lfloor \tau\fs \rfloor+1}{\fs} ), & |i-j| = \lfloor \tau\fs \rfloor; \\
        \frac{1}{2\pi} L (\frac{\lfloor \tau\fs \rfloor}{\fs}-\tau), & |i-j| = \lfloor \tau\fs \rfloor+1; \\
        0, & \text{otherwise},
    \end{cases}
    \label{eq:Sigma_d_ij}
\end{equation}
where $p>0$ and $q<0$. 
Earlier in the discussion of the likelihood model, we set $p=\hat{h}(\SNRetaHat)$.
When $\SNReta$ is high (above 10~dB), $q \approx -\Frac{p}{2}$.
Calculation of $\zeta_n$ is effectively a first difference of the phase, hence there are negative correlations between adjacent samples.
When $\SNReta$ is low (below 10~dB), $q$ becomes larger and the negative correlations  diminish, as shown in \cref{fig:snrvar}.
We also mark in \cref{fig:snrvar} the range of $\SNReta$ values occurring in the simulation results for \cref{fig:dest,fig:dvest}. 

\begin{figure}
    \centering
    \includegraphics[width=0.99\linewidth]{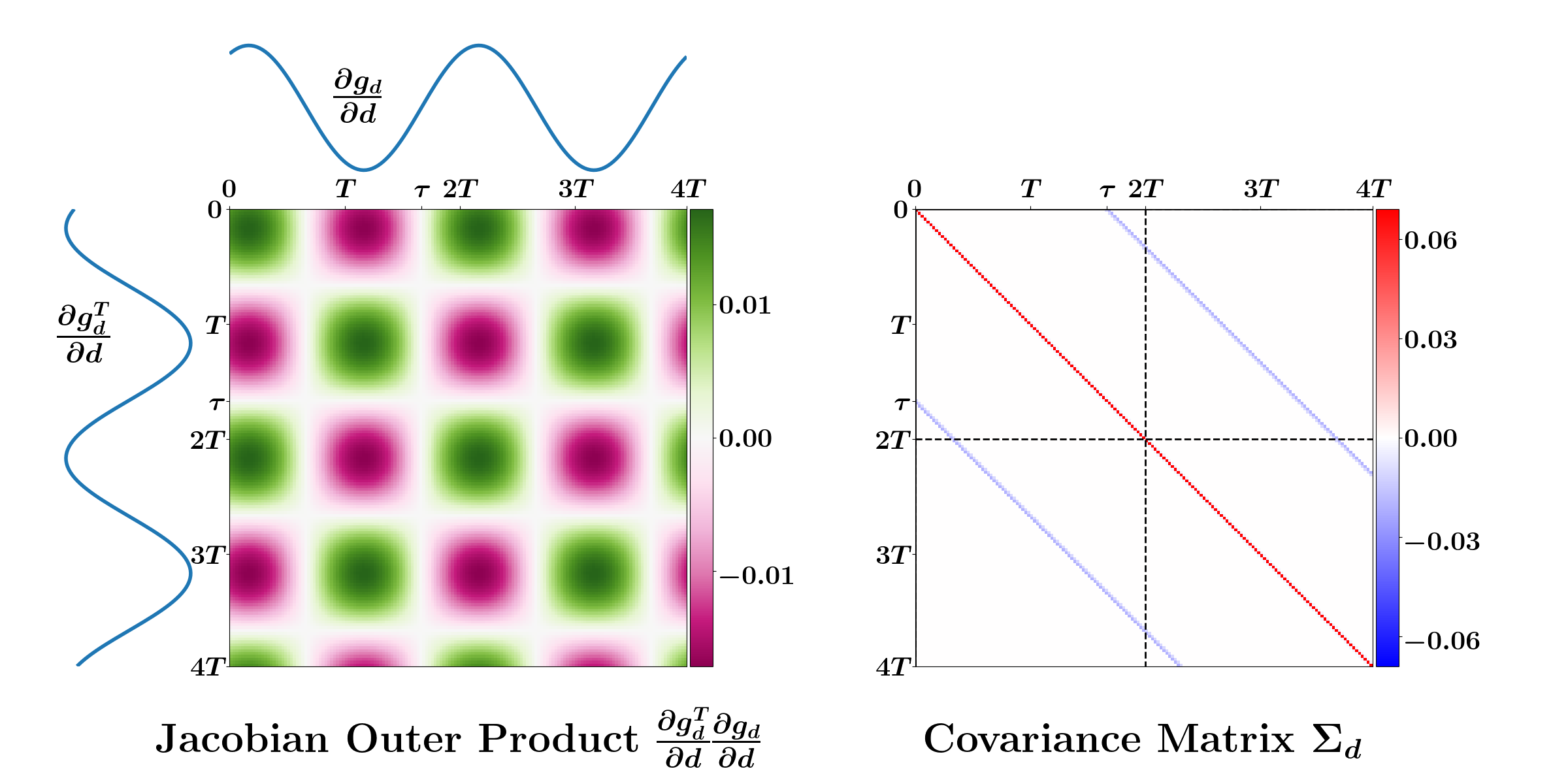}
    \caption{Illustration of Jacobian outer product (shown example is using sinusoidal modulation) and covariance matrix of noise for some distance with corresponding delay $T < \tau < 2T$. Samples are taken across $4T$ which is 2 periods. One dashed square represent a covariance matrix for one period. Red indicate positive values and blue indicate negative values (zero otherwise). Off-diagonal negative correlations will be shifted according to the delay.
    The rate of improvement in RMSE comparing to observing 1 period will vary for different distances due to correlations. In our case, estimations for long distances ($\tau>T$) will improve more than $\sqrt{2}$ times since the number of opposite signs increase more than the number of same signs. }
    \label{fig:cov}
\end{figure}

In the case of a simple sample mean estimation, having negative correlations will improve the accuracy.
In our case, correlation between the $i$th and $j$th samples will improve the accuracy when
\begin{equation}
    \text{sign}\!\left(\left[\frac{\diff \bm g_d^T}{\diff d}\frac{\diff \bm g_d}{\diff d}\right]_{ij}\right)
    \not = \text{sign}\!\left(\left[ \bm \Sigma_d\right]_{ij}\right).
    \label{eqn:jaccovsign}
\end{equation}
In \cref{fig:cov}, we show an example of both the Jacobian outer product and the covariance matrix.
We observe that for longer ranges ($\tau > T$), the number of opposite signs increases more than the same signs when the correlations exist across periods.
The signs of elements in the Jacobian outer product at $|i-j|=1$ tend to be positive. 
Hence at low $\SNReta$, the performance will drop significantly due to loss of negative correlations. 
The third and fourth cases in  \eqref{eq:Sigma_d_ij} are also negative values, which sum to
$- L\Delta_t / (2\pi)$.

\subsection{Comparing with Numerical Results}

\begin{figure}
    \centering \includegraphics[width=0.8\linewidth]{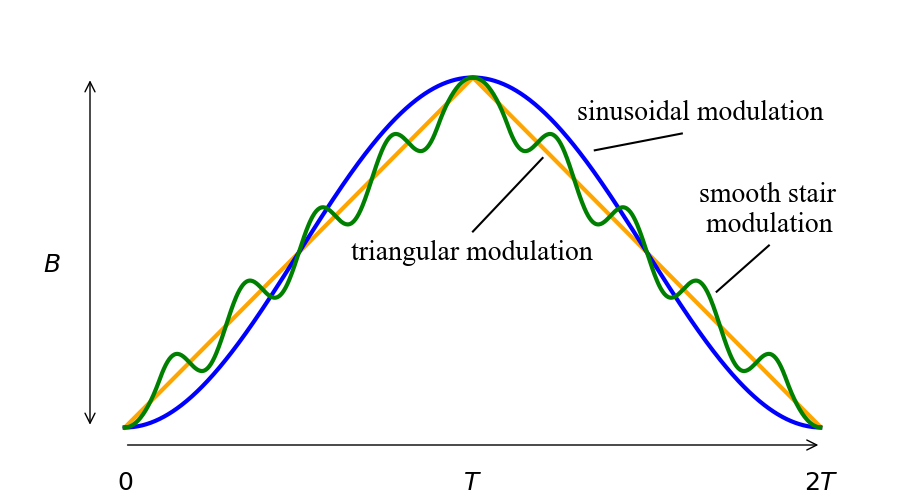}
    \caption{Modulation functions $a(t)$ for triangular (orange), sinusoidal (blue), and smooth stair modulation (green). 
    The smooth stair function is a continuously differentiable function that we have defined for the purpose of validating the misspecified CRB\@. }
    \label{fig:smoothstair}
\end{figure}

The MCRB equation \eqref{eqn:mcrb} indicates that our estimator accuracy increases when the IF $\gwtf$ changes more rapidly. 
To illustrate faster changes in frequency without decreasing chirp duration $T$ or increasing bandwidth $B$,
we introduce the
``smooth stair'' modulation function,
shown in \cref{fig:smoothstair}.
The analytical form for instantaneous frequency of smooth stair modulation is
\begin{equation}
\begin{split}
    &a_{\text{smoothstair}}(t) =\\
    &\begin{cases}
        c_1 t^2, & t \in [0, t_0]; \\
        \frac{B}{T} t - c_2\sin(c_3(t-T/2)),
        & t \in [t_0, T-t_0]; \\
        B-c_1 (t-T)^2, & t \in [T-t_0, T+t_0]; \\
        \frac{B}{T} (2T-t)- c_2\sin(c_3(3T/2-t)),
        & t \in [T+t_0, 2T-t_0]; \\
        c_1 (t-2T)^2, & t \in [2T-t_0, 2T], \\
    \end{cases}
    \label{eq:smoothstair}
\end{split}
\end{equation}
where $t_0=T/10$, $c_1=13.68B$, $c_2=-0.069B$, and $c_3=30/T$.
The values of $c_1$ and $c_2$ were chosen to make the function  continuously differentiable for the given $t_0$ and $c_3$.
The smooth stair or similar modulation could be achieved by adapting the iterative learning pre-distortion approach of Zhang et al.~\cite{zhang_laser_2019}.
The feasibility of generating such modulation will depend on the hardware's ability to achieve fast chirp rates without introducing systematic error due to carrier and thermal effects~\cite{pinto_enhanced_w_tunable_2025}.
A more detailed discussion on the hardware implementation is outside the scope of this paper.

\begin{figure*}
    \centering\includegraphics[width=0.99\linewidth]{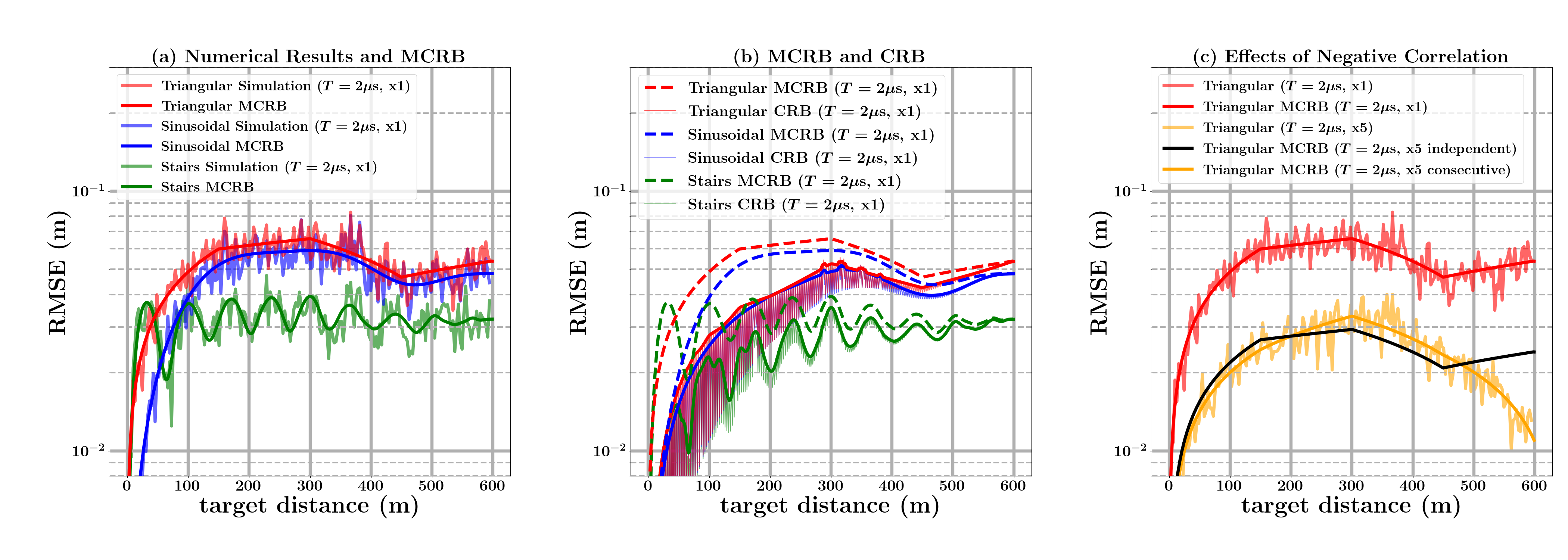}
    \caption{
    MCRB comparisons. 
    In (a), we compare the theoretical (MCRB) and numerical performance of our IFF method for triangular, sinusoidal, and ``smooth stair'' modulations. 
    The MCRB closely matches the simulation results, predicting the superior performance of the smooth stair modulation. 
    In (b), the CRB oscillates rapidly as a function of distance, so we also plot its upper envelope using a local maximum filter.
    As expected, the CRB is a lower bound on the MCRB as it quantifies the optimal performance without mismatch.
    In (c), we demonstrate how negative correlations in $\Sigma_d$ over longer acquisition times yields lower error than averaging the results of multiple shorter acquisitions.
    }
    \label{fig:mcrb}
\end{figure*}

While we make no optimality claims for smooth stair modulation, MCRB analysis suggests improved performance, and Monte Carlo simulations confirm this.
In \cref{fig:mcrb}(a), we show MCRBs along with Monte Carlo simulation results to show the consistency between the two.
The smooth stair modulation has oscillating error but better overall accuracy compared to the other two modulations,  consistent with our hypothesis that large derivatives in $a(t)$ lead to better overall accuracy. 
In \cref{fig:mcrb}(b), we compare the MCRB and CRB for each modulation function. 
The gap between the bounds is caused by the lack of consideration of correlation in $\LIFF$. 
Triangular modulation suffers the largest penalty from this model mismatch.

In \cref{fig:mcrb}(c), we show the effects of negative correlations in the estimations. 
When estimation is performed using 5 periods instead of 1 period, we expect averaging of estimates from each period would cause the RMSE to improve by $\sqrt{5}$ times.
However, these five periods are consecutive
rather than five independent trials of an experiment.
Consecutive periods can introduce more negative correlations between the samples, which improves the RMSE by more than by a factor of $\sqrt{5}$.

\subsection{Approximating Total Error}

Although the MCRB accurately predicts the RMSE for a particular distance $d$, a practitioner may be more interested in the average performance for a particular modulation function $a(t)$ over the entire unambiguous distance range of zero to $cT$.
We introduce the notion of mean MCRB (MMCRB) 
\begin{equation}
    \text{MMCRB} = \frac{1}{cT}\int_0^{cT} \text{MCRB}_d \diff d
    \label{eqn:mmcrb_exact}
\end{equation}
to quantify the overall performance of our estimator without needing to perform extensive Monte Carlo simulations.
Still, comparing or optimizing modulation functions based on~\eqref{eqn:mmcrb_exact} is computationally expensive.
Using the relation $\left.\frac{\partial \bm{g}_d}{\partial d}\right|_{d=0}=\frac{2}{c}\cdot\frac{\diff \bm a}{\diff t}$
and treating
$\frac{\partial \tilde{\bm{g}}_d}{\partial d}$
as constant in the integral \eqref{eqn:mmcrb_exact}, we can approximate MMCRB by
\begin{gather}
    \text{MMCRB}\approx \frac{1}{cT}\cdot \frac{4}{c^2}\cdot\frac{ \frac{\diff \bm a^T }{\diff t} \left( \int_0^{cT} \Sigma_d \diff d\right)\frac{\diff \bm a}{\diff t} }{\left( \frac{\diff \bm a^T}{\diff t}\frac{\diff \bm a}{\diff t}\right)^2}. 
    \label{eqn:MMCRB}
\end{gather}
This approximation requires only one vector--matrix--vector product. 
Moreover, the above formulation of the MMCRB implies that a larger deviation in IF (i.e. $\frac{\diff \bm a^T}{\diff t}\frac{\diff \bm a}{\diff t}$) leads to lower overall distance estimation error.

\begin{table}
    \centering
    \caption{Root Mean MCRB ($10^{-2}\cdot{\rm m}$) }
    \label{tab:MMCRB}
    \begin{tabular}{  lccc  } 
        \toprule
        & Triangle & Sinusoidal & Smooth Stair \\
        \midrule
        MMCRB~\eqref{eqn:mmcrb_exact} & $5.362$ & $4.820$ & $3.214$ \\ 
        Approximation~\eqref{eqn:MMCRB} & $5.355$ & $4.815$ & $3.209$ \\
        \bottomrule
    \end{tabular}
    
    \vspace{2pt}
    \begin{minipage}{\linewidth}
        \footnotesize
        \textit{Note:} The MMCRB quantifies the average distance accuracy
        over the unambiguous range, showing good overall
        performance
        for the smooth stair modulation. 
        The approximation \eqref{eqn:MMCRB} is a computationally efficient alternative to
        \eqref{eqn:mmcrb_exact}.
    \end{minipage}
\end{table}

\Cref{tab:MMCRB} validates our  MMCRB approximation from~\eqref{eqn:MMCRB}. 
For all three modulations, we compare the numerical integration of MCRB over distances from \SI{1}{\meter} to \SI{600}{\meter} versus the MMCRB approximation formula. 
It also makes clear that the smooth stair modulation 
leads to a lower overall RMSE.

\section{Conclusion}\label{sec:conclusion}

We presented the limitations of conventional FMCW processing,  discussing the unambiguous space of distance/velocity from a pair of constant beat frequencies generated from triangular modulation. 
We propose
a novel framework that instead models the entire signal
through the 
instantaneous phase or frequency. 
Modeling beyond the constant beat frequencies helps expand the unambiguous distance/velocity space, increases the sample efficiency, and removes the dependence on linear frequency modulation, 
which we demonstrate through simulations.
When using our approach, there is no longer a need to compromise between the sensitivity of measurements versus the maximum distance because of the sampling rate.

We proposed two parameter estimation approaches with different tradeoffs.
Matched filtering is conceptually and algorithmically straightforward, and it is fast for distance-only estimation; however, the performance degrades with distance due to phase noise---especially for triangular modulation---and the computational complexity scales poorly for distance--velocity estimation.
Our instantaneous frequency fitting approach requires more careful noise modeling and parameter selection, but the gradient-based method has excellent performance across different modulation functions, distances, and velocities.

Our proposed approaches hinges on three key assumptions:
a known modulation function,
a single echo, and complex-valued measurements.
Future research may explore the opportunities and challenges of breaking these assumptions for more practical lidar systems.
Knowing the modulation function $a(t)$ presumes a previous characterization step~\cite{ahn_analysis_2007}, which has been demonstrated for tunable lasers~\cite{siddharth_ultrafast_2025}.
Further work should investigate the robustness of such a calibration, e.g., to environmental changes such as temperature, or whether an additional interferometer reference branch would be necessary.
Restricting our attention to a single reflector allows the single received instantaneous frequency trajectory to be estimated reasonably well through phase differentiation.
Extending the IFF method to settings with multiple echoes would require extracting the IFs for multiple components \cite{Boashash:92b, caromna_multiridge_1999,legros_multiif_sparse_2024,Li_ifsynchro_2024,chen_crossoverif_2023}, 
which becomes difficult if those IFs are overlapped (as they would likely be if aliased).
The matched filter approach may be more suitable for multiple echoes, as it avoids the IF extraction step.
Finally, the IFF could be extended to real-valued signals using the Hilbert transform, although the phase noise statistics would deviate from our model assumption and likely degrade performance. 
Other time-frequency estimation methods such as synchrosqueezing~\cite{li_synchroridge_2021,zheng_fmcwcorrecmss_2023} or the superlet transform~\cite{moca_time-frequency_2021, feng_nonlinear_2025} could potentially provide more robustness.

While our focus is on parameter estimation from FMCW lidar signals, many of the core ideas of our work could generalize to more complicated estimation problems, such as regression of higher dimensional parametric models on wrapped observed signals as in phase unwrapping and unlimited sampling~\cite{bhandari_unlimitedsample_2021}. 
The non-convexity introduced by wrapping creates additional challenges to optimization and, for higher dimensional problems, computational complexity must be carefully considered. 
We hope that our simpler scenario can bring new insights. 
Further work may also connect our example through such abstraction to other powerful and well-established fields and theories such as information geometry~\cite{amari_infogeo_2016} to create more tools for estimation and inference on non-standard topology.

\appendices

\crefalias{section}{appendix}

\begin{figure*}[!th]

\begin{equation}
\bm J = \frac{1}{\sigma^2 \fs}\begin{bmatrix}
2N  & 0 & 0 & 0\\
0 & 2\alpha^2 N \fs & 2\alpha^2\pi N(2N-1) 
& -2\alpha^2 \pi\gamma N^2 \\
0 & 2\alpha^2\pi N(2N-1) & \Frac{4\alpha^2 \pi^2N(2N-1)(4N-1)}{3} & -4\alpha^2 \pi^2\gamma N^2(2N-1) \\
0 & - 2\alpha^2 \pi \gamma N^2 & -4\alpha^2 \pi^2\gamma N^2(2N-1) &  \Frac{4\alpha^2 \pi^2\gamma^2N(2N-1)(4N-1)}{3}
\end{bmatrix}\label{eq:FIM}
\end{equation}
\vspace{1ex}
\hrule

\end{figure*}

\section{Cram\'er--Rao Bound for Ideal CBF Signals with AWGN}
\label{apx:crb_cbf}

To better understand the impact of the bandwidth $B$ and chirp duration $T$, we consider an idealized measurement scenario with triangular modulation that assumes only additive white Gaussian noise and no non-CBF region. 
We define our signal model $s(t_n)$ as
$$
s(t_n) = \begin{cases}
\alpha\exp(j2\pi(\gamma\tau + f) t_n + jb), & 0 \leq n <N; \\
\alpha\exp(j2\pi(-\gamma\tau + f) t_n + jb), & N \leq n <2N,
\end{cases}
$$
where $\gamma = \Frac{B}{T}$ is the chirp rate of the triangular modulation. 
Our measured signal is
$$
u(t_n) = s(t_n) + w(t_n),
$$
where $w(t_i)$ is circularly-symmetric, complex white Gaussian noise with real and imaginary components that are independent and zero-mean with variance $\sigma^2$. 
Let $t_0 = 0$.
The Fisher information is
$$
[\bm J]_{ij} = \frac{1}{\sigma^2}\mathrm{Re}\left\{\sum_{n=0}^{2N-1}\frac{\partial \bar{s}(t_n)}{\partial \theta_i} \frac{\partial s(t_n)}{\partial \theta_j} \right\}.
$$
We let $\theta_1 = \alpha$, $\theta_2=b$, $\theta_3 = f$,
and $\theta_4=\tau$. 
The derivatives with respect to each parameters are
\begin{align}
\frac{\partial s(t_n)}{\partial \alpha} &= \frac{1}{\alpha}s(t_n), \quad \frac{\partial s(t_n)}{\partial b} = js(t_n), \nonumber \\
\frac{\partial s(t_n)}{\partial f} &= j2\pi t_n s(t_n), \nonumber \\
\frac{\partial s(t_n)}{\partial \tau} &= \begin{cases}
j2\pi\gamma t_n s(t_n), & 0 \leq n < N; \\
-j2\pi\gamma t_n s(t_n), & N \leq n < 2N.
\end{cases} \nonumber
\end{align}
The Fisher information matrix $\bm{J}$ is then given in~\eqref{eq:FIM}.
Taking its inverse, we see the lower bounds on the variance for unbiased estimates $\hat{\tau}$ and $\hat{f}$ are
\begin{align}
        \Var(\hat{\tau}) &\geq \frac{\sigma^2 3 \fs^{2} \left(2 N + 1\right)}{4 N \gamma^{2} \pi^{2} a^{2} \left(N - 1\right) \left(N + 1\right) \left(4 N - 1\right)} \nonumber \\
        &\propto \frac{\fs^2N}{\gamma^2 N^4} = \frac{1}{B^2T\fs}
\end{align}
and
\begin{align}
        \Var(\hat{f}) &\geq \frac{\sigma^2 3 \fs^{2} \left(13 N^{2} - 12 N + 2\right)}{4 N \pi^{2} a^{2} \left(N - 1\right) \left(N + 1\right) \left(2 N - 1\right) \left(4 N - 1\right)} \nonumber \\
        &\propto \frac{\fs^2N^2}{N^5} = \frac{1}{T^3\fs},
\end{align}
where we substituted $N\approx T\fs$.

\section{Cram\'er--Rao Bound for Delay Estimation}
\label{apx:crb_delay}

In addition to the MCRB, which we mainly use for the theoretical analysis, we also derive the traditional Cram\'er-Rao bound (CRB). 
The major difference from the MCRB is that this CRB assumes we are leveraging the sample correlation to estimate the parameters.
In deriving the CRB, we assume we are fitting the unwrapped and normalized IF $\tilde{\bm g}_d = (\Frac{1}{\fs})\bm g_d$ to the estimated IF $\bm \zeta$ corrupted with additive Gaussian noise. 
We will assume the Doppler shift is known to be zero and
$$
\bm \zeta\sim \mathcal{N}(\tilde{\bm g}_d, \bm \Sigma_d).
$$
Note the covariance matrix depends on the distance $d$. 
Hence the log-likelihood is 
$$
\log p_d(\bm \zeta) \propto -\frac{1}{2}\log |\bm \Sigma_d| - \frac{1}{2}(\bm \zeta-\tilde{\bm g}_d)^T\bm \Sigma^{-1}_d (\bm \zeta-\tilde{\bm g}_d).
$$
Taking the derivative of $\log p_d(\bm \zeta)$ gives us
\begin{align*}
\frac{\partial}{\partial d}\log p_d(\bm \zeta)
&= -\frac{1}{2}\frac{\partial}{\partial d}\log |\bm \Sigma_d| %
+\frac{\partial\tilde{\bm g}_d^T}{\partial d} \bm \Sigma^{-1}_d (\bm \zeta-\tilde{\bm g}_d) \\
&\qquad + \frac{1}{2}(\bm \zeta - \tilde{\bm g}_d)^T \frac{\partial\bm \Sigma^{-1}_d}{\partial d} (\bm \zeta-\tilde{\bm g}_d),
\end{align*}
where $\frac{\partial\bm \Sigma^{-1}_d}{\partial d}$ is an entry-wise derivative. 
By Jacobi's formula,
\begin{equation*}
    \begin{split}
        \frac{\partial}{\partial d}\log|\bm \Sigma_d|
        &= \frac{\frac{\partial}{\partial d}|\bm \Sigma_d|}{|\bm \Sigma_d|} \\
        &= \frac{|\bm \Sigma_d|\cdot \tr\!\left( \frac{\partial \bm\Sigma_d}{\partial d}  \bm\Sigma_\tau^{-1}\right)}{|\bm \Sigma_d|}\\
        &= \tr\!\left( \frac{\partial \bm\Sigma_d}{\partial d}  \bm\Sigma_d^{-1}\right),
    \end{split}
\end{equation*}
and the derivative of the matrix inverse can be written as
\begin{equation*}
\frac{\partial\bm \Sigma^{-1}_d}{\partial d} = -\bm\Sigma^{-1}\frac{\partial\bm \Sigma_d}{\partial d} \bm\Sigma^{-1}.
\end{equation*}

It should be noted that $\bm \Sigma^{-1}_d$ and $\bm g_d$ may not be differentiable everywhere. 
The Fisher information is the expectation of $\frac{\partial}{\partial d}\log p_d(\bm \zeta)^2$,
which is
\begin{equation*}
    \begin{split}
        J_d &= \mathbb{E}\biggl [ \frac{1}{4} \tr\!\left( \frac{\partial \bm\Sigma_d}{\partial d}  \bm\Sigma_d^{-1}\right)^2\\
        & \hspace{0.2em} + \frac{\partial\tilde{\bm g}_d^T}{\partial d} \bm \Sigma^{-1}_d (\bm \zeta-\tilde{\bm g}_d) (\bm \zeta-\tilde{\bm g}_d)^T   \bm \Sigma^{-1}_d  \frac{\partial\bm g_d}{\partial d}\\
        & \hspace{0.2em} +\frac{1}{4}(\bm \zeta - \tilde{\bm g}_d)^T \frac{\partial\bm \Sigma^{-1}_d}{\partial d} (\bm \zeta-\tilde{\bm g}_d) (\bm \zeta - \tilde{\bm g}_d)^T \frac{\partial\bm \Sigma^{-1}_d}{\partial d} (\bm \zeta-\tilde{\bm g}_d)\\
        & \hspace{0.2em} - \tr\!\left( \frac{\partial \bm\Sigma_d}{\partial d}  \bm\Sigma_d^{-1}\right) (\bm \zeta-\tilde{\bm g}_d)^T   \bm \Sigma^{-1}_d  \frac{\partial\tilde{\bm g}_d}{\partial d}\\
        & \hspace{0.2em} -\frac{1}{2} \tr\!\left( \frac{\partial \bm\Sigma_d}{\partial d}  \bm\Sigma_d^{-1}\right) (\bm \zeta-\tilde{\bm g}_d)^T   \frac{\partial \bm\Sigma_d^{-1}}{\partial d}  (\bm \zeta-\tilde{\bm g}_d) \\
        & \hspace{0.2em} + \frac{\partial\tilde{\bm g}_d^T}{\partial d} \bm \Sigma^{-1}_d (\bm \zeta-\tilde{\bm g}_d) (\bm \zeta-\tilde{\bm g}_d)^T   \frac{\partial \bm\Sigma_d^{-1}}{\partial d}  (\bm \zeta-\tilde{\bm g}_d) \biggl ].
    \end{split}
\end{equation*}
The first term is independent of the random variable $\bm \zeta$.
The second can be easily calculated since $\mathbb{E}[(\bm \zeta-\tilde{\bm g}_d) (\bm \zeta-\tilde{\bm g}_d)^T] = \bm \Sigma_d$.
The fourth and sixth terms are zero since they are odd moments of a zero-mean Gaussian.
The third and the fifth terms can be calculated using Isserlis' theorem as
\begin{equation*}
\begin{split}
\mathbb{E}\biggl[&(\bm \zeta - \tilde{\bm g}_d)^T \frac{\partial\bm \Sigma^{-1}_d}{\partial d} (\bm \zeta-\tilde{\bm g}_d) (\bm \zeta - \tilde{\bm g}_d)^T \frac{\partial\bm \Sigma^{-1}_d}{\partial d} (\bm \zeta-\tilde{\bm g}_d)\biggl] \\
&= \tr\!\left( \frac{\partial\bm \Sigma^{-1}_d}{\partial d} \bm\Sigma_d\right)^2 + 2\cdot \tr\!\left( \frac{\partial\bm \Sigma^{-1}_d}{\partial d} \bm\Sigma_d\frac{\partial\bm \Sigma^{-1}_d}{\partial d} \bm\Sigma_d\right) \\
&= \tr\!\left( \frac{\partial \bm\Sigma_d}{\partial d} \bm \Sigma_d^{-1} \right)^2 + 2\cdot \tr\!\left( \frac{\partial \bm\Sigma_d}{\partial d} \bm \Sigma_d^{-1}  \frac{\partial \bm\Sigma_d}{\partial d} \bm \Sigma_d^{-1} \right)
\end{split}
\end{equation*}
and
\begin{align*}
\mathbb{E}[ (\bm \zeta-\tilde{\bm g}_d)^T \frac{\partial \bm\Sigma_d^{-1}}{\partial d} (\bm \zeta-\tilde{\bm g}_d) ]
&= \tr\!\left( \frac{\partial \bm\Sigma_d^{-1}}{\partial d}  \bm \Sigma_d \right) \\
&= \tr\!\left( \frac{\partial \bm\Sigma_d}{\partial d}  \bm \Sigma_d^{-1} \right),
\end{align*}
which yields
\begin{equation*}
    \begin{split}
        J_d &= \frac{1}{4} \tr\!\left( \frac{\partial \bm \Sigma_d}{\partial d} \bm \Sigma_d^{-1}\right)^2  \\
        &\qquad + \left( \frac{\partial \tilde{\bm g}_d}{\partial \tau}\right)^T \bm \Sigma_d^{-1} \left( \frac{\partial \tilde{\bm g}_d}{\partial d}\right) \\
        &\qquad + \frac{1}{4} \tr\!\left( \frac{\partial \bm\Sigma_d}{\partial d} \bm \Sigma_d^{-1} \right)^2 + \frac{1}{2} \tr\!\left( \frac{\partial \bm\Sigma_d}{\partial d} \bm \Sigma_d^{-1}  \frac{\partial \bm\Sigma_d}{\partial d} \bm \Sigma_d^{-1} \right)\\
        & \qquad + 0 \\
        & \qquad - \frac{1}{2} \tr\!\left( \frac{\partial \bm \Sigma_d}{\partial d} \bm \Sigma_d^{-1}\right)^2\\
        & \qquad + 0 \\
        &= \left( \frac{\partial \tilde{\bm g}_d}{\partial d}\right)^T \bm \Sigma_d^{-1} \left( \frac{\partial \tilde{\bm g}_d}{\partial d}\right) + \frac{1}{2} \tr\!\left( \frac{\partial \bm\Sigma_d}{\partial d} \bm \Sigma_d^{-1}  \frac{\partial \bm\Sigma_d}{\partial d} \bm \Sigma_d^{-1} \right).
    \end{split}
\end{equation*}
The Cram\'er--Rao bound is simply the inverse of the above expression.

\section{Derivation of Misspecified Cram\'er--Rao Bound for Delay Estimation}
\label{apx:mcrb_delay}

The misspecified Cram\'er--Rao bound, 
the lower bound on the estimator $\hat{\theta}$, is calculated by 
\begin{equation*}
\Var(\hat{\theta}~|~\theta) \geq \frac{\beta(\theta)}{\alpha^2(\theta)},
\end{equation*}
where
\begin{align*}
\alpha(\theta) &= \mathbb{E}_{p}\biggl [ \frac{\partial^2}{\partial \theta^2}\log q_\theta(x) \biggl ], \\
\beta(\theta) &= \mathbb{E}_{p}\biggl [ \left(\frac{\partial}{\partial \theta}\log q_\theta(x) \right)^2 \biggr].
\end{align*}
The estimator is assumed to MS-unbiased~\cite{vuong_mcrb_1986, fortunati_mcrb_2017}, i.e.,
\begin{equation*}
    \mathbb{E}[\hat{\theta}] = \theta = \underset{\theta}{\text{argmin }} D_{KL}(p||q_\theta),
\end{equation*}
where $p$ is the true distribution and $q_\theta$ is the approximating distribution function. 
In order to compute the MCRB, we write $p$ as Gaussian and $q_d$ as a Gaussian with diagonal covariance matrix:
$$
p_d = \mathcal{N}(\tilde{\bm g}_d, \bm \Sigma_d), \quad q_d = \mathcal{N}(\tilde{\bm g}_d, \sigma^2\bm I),
$$
where $q_d$ is the misspecified likelihood used for our estimator. Since the Gaussian distribution is symmetric, MS-unbiased implies unbiased. Hence, the MCRB is the lower bound on the MSE
\begin{equation*}
\mathbb{E}[(\hat{d}-d)^2~|~d] \geq \frac{\beta(d)}{\alpha^2(d)}.
\end{equation*}
When the misspecified likelihood model is a least-squares estimator, the likelihood is a normal distribution with a diagonal covariance matrix, where the log-likelihood is
\begin{equation*}
\log q_d(\bm\zeta) \propto -\frac{1}{2\sigma^2}(\bm\zeta-\tilde{\bm g}_d)^T (\bm \zeta - \tilde{\bm g}_d).
\end{equation*}

The quantities $\alpha(d)$ and $\beta(d)$ are obtained as follows:
\begin{align*}
 \alpha(d) &= \mathbb{E}\biggl [ \frac{1}{2\sigma^2}\frac{\partial}{\partial d} \frac{\partial \tilde{\bm g}_d}{\partial d}(\bm\zeta-\tilde{\bm g}_d)\biggl ] \nonumber \\
 &= \mathbb{E}\biggl [ \frac{1}{2\sigma^2} \frac{\partial^2 \tilde{\bm g}_d}{\partial d^2}(\bm\zeta-\tilde{\bm g}_d) - \frac{1}{2\sigma^2}\frac{\partial \tilde{\bm g}_d^T}{\partial d}\frac{\partial \tilde{\bm g}_d}{\partial d}\biggl ] \nonumber \\
 &= - \frac{1}{2\sigma^2}\frac{\partial \tilde{\bm g}_d^T}{\partial d}\frac{\partial \tilde{\bm g}_d}{\partial d},
\end{align*}
\begin{equation*}
 \beta(d) = \mathbb{E}\biggl [ \left( \frac{1}{2\sigma^2}\frac{\partial \bm g^T_d}{\partial d} (\bm \zeta - \tilde{\bm g}_d )\right)^2 \biggl ] = \frac{1}{4\sigma^4}\frac{\partial \tilde{\bm g}_d}{\partial d} \bm \Sigma_d \frac{\partial \tilde{\bm g}_d}{\partial d}.
\end{equation*}
Hence the MCRB is
\begin{equation*}
\Var(\hat{d}~|~d) \geq \frac{\frac{\partial \tilde{\bm g}_d}{\partial d} \bm \Sigma_d \frac{\partial \tilde{\bm g}_d}{\partial d}}{\left(\frac{\partial \tilde{\bm g}_d}{\partial d}  \frac{\partial \tilde{\bm g}_d}{\partial d}\right)^2}.
\end{equation*}

\bibliographystyle{ieeetr}
\bibliography{ref}

\end{document}